\documentclass[aps,prx,twocolumn,superscriptaddress,showpacs,showkeys,
notitlepage]{revtex4-1}

\usepackage{graphicx}
\usepackage{dcolumn}
\usepackage{bm}
\usepackage{float}
\usepackage{amsmath}

\begin{document}

\title{Increased lifetime of metastable skyrmions by controlled doping}

\author{M. T. Birch}
\address{Durham University, Centre for Materials Physics, Durham, DH1 3LE, United Kingdom}
\address{Diamond Light Source, Didcot, OX11 0DE, United Kingdom}
\author{R. Takagi}
\address{RIKEN Center for Emergent Matter Science (CEMS), Wako, 351-0198, Japan}
\author{S. Seki}
\address{RIKEN Center for Emergent Matter Science (CEMS), Wako, 351-0198, Japan}
\author{M. N. Wilson}
\address{Durham University, Centre for Materials Physics, Durham, DH1 3LE, United Kingdom}
\author{F. Kagawa}
\address{RIKEN Center for Emergent Matter Science (CEMS), Wako, 351-0198, Japan}
\address{University of Tokyo, Department of Applied Physics, Bunkyo-ku 113-8656, Japan}
\author{A. \v{S}tefan\v{c}i\v{c}}
\address{University of Warwick, Department of Physics, Coventry, CV4 7AL, United Kingdom}
\author{G. Balakrishnan}
\address{University of Warwick, Department of Physics, Coventry, CV4 7AL, United Kingdom}
\author{R. Fan}
\address{Diamond Light Source, Didcot, OX11 0DE, United Kingdom}
\author{P. Steadman}
\address{Diamond Light Source, Didcot, OX11 0DE, United Kingdom}
\author{C. J. Ottley}
\address{Durham University, Earth Sciences, Durham, DH1 3LE, United Kingdom}
\author{M. Crisanti}
\address{University of Warwick, Department of Physics, Coventry, CV4 7AL, United Kingdom}
\address{Institut Laue-Langevin, CS 20156, 38042 Grenoble Cedex 9, France}
\author{R. Cubitt}
\address{Institut Laue-Langevin, CS 20156, 38042 Grenoble Cedex 9, France}
\author{T. Lancaster}
\address{Durham University, Centre for Materials Physics, Durham, DH1 3LE, United Kingdom}
\author{Y. Tokura}
\address{RIKEN Center for Emergent Matter Science (CEMS), Wako, 351-0198, Japan}
\address{University of Tokyo, Department of Applied Physics, Bunkyo-ku 113-8656, Japan}
\author{P. D. Hatton}
\address{Durham University, Centre for Materials Physics, Durham, DH1 3LE, United Kingdom}

\begin{abstract}
Previous observations of metastable magnetic skyrmions have shown that close to the equilibrium pocket the metastable state has a short lifetime, and therefore rapid cooling is required to generate a significant skyrmion population at low temperatures. Here, we report that the lifetime of metastable skyrmions in Cu$_2$OSeO$_3$ is extended by a factor of 50 with the introduction of only 2.5\% zinc doping, allowing over 50\% of the population to survive when field-cooling at a rate of just 1 K/min. Our systematic study suggests that the lifetime enhancement is due to the removal of spins by the non-magnetic dopant, which entropically limits the number of skyrmion decay pathways. We expect that doping can be exploited to control the lifetime of the metastable SkL state in other chiral magnets, offering a method of engineering skyrmion materials towards application in future devices.
\end{abstract}

\maketitle
\section{Introduction}
The magnetic skyrmion lattice (SkL) is an example of a state of matter protected by a topological energy barrier. This barrier arises from the inability to continuously transform between spin structures with different topological charge, $N$. \cite{nagaosa_topological_2013}. This number effectively counts the number of times the spin direction wraps around a unit sphere, yielding $N$ = 1 for the skyrmion state, and $N$ = 0 for the competing, topologically trivial, helical, conical and spin polarised magnetic states. The resulting energy barrier stabilises the skyrmion state, protecting it from destruction by small perturbations that cause continuous deformations in the spin structure, but not from discontinuous changes, as may be caused by large thermal fluctuations. 

\begin{figure}
\centering
\includegraphics[width=0.41\textwidth]{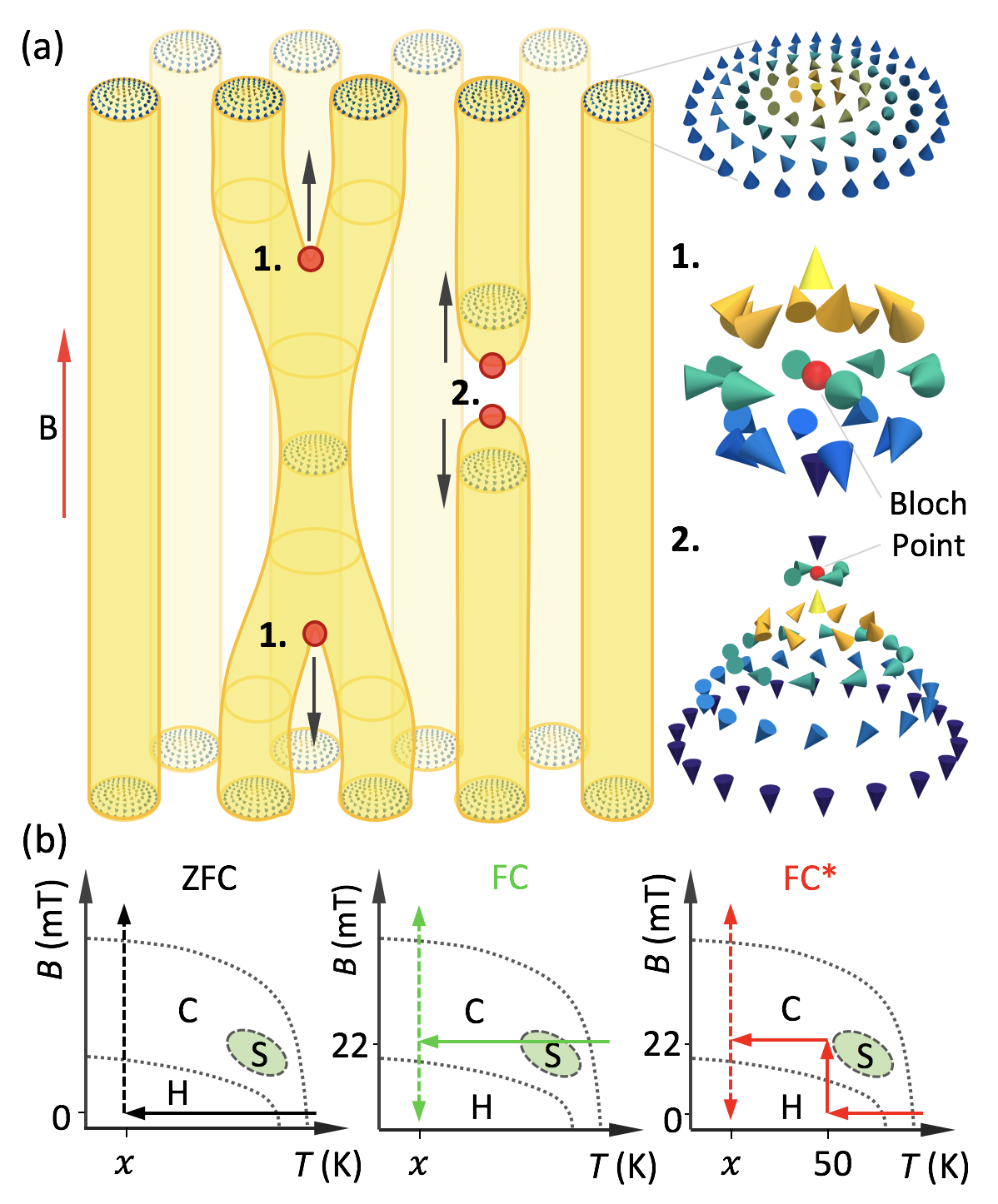}
\caption{(a) Illustration of the vertical, string-like structure of a hexagonal lattice of Bloch skyrmions. Two scenarios for the Bloch point unwinding mechanism are displayed. (b) Schematic diagrams of the zero field-cooling (ZFC) (black) and field-cooling (FC, FC*) (green, red) measurement procedures. Solid lines indicate the cooling procedure, while dashed lines indicate the measurement path. Dotted lines illustrate the boundaries between the helical (H), conical (C), and skyrmion (S) phases.}
\label{fig1}
\end{figure}

In bulk materials, skyrmions are typically observed in chiral systems such as the B20 structures MnSi \cite{muhlbauer_skyrmion_2009}, Fe$_{0.5}$Co$_{0.5}$Si \cite{yu_real-space_2010} and FeGe \cite{yu_near_2011} (space group $P2_{1}3$), where the broken inversion symmetry results in an antisymmetric Dzyaloshinskii-Moriya Interaction (DMI) which favours canting between neighbouring spins. The competition of this interaction with the ferromagnetic exchange and Zeeman interactions produces noncolinear spin textures such as the helical and conical states. In the presence of thermal fluctuations near $T_{\rm{c}}$ \cite{rosler_spontaneous_2006}, at particular values of crystal anisotropies \cite{chacon_observation_2018}, or with the addition of magnetic frustration \cite{karube_disordered_2018}, the magnetic SkL state is energetically favoured, forming a skyrmion lattice in a plane perpendicular to the direction of an applied magnetic field. In bulk materials, this lattice has an extended, string-like structure in the field direction, as illustrated in Fig. \ref{fig1}(a) \cite{yokouchi_current-induced_2018}. Other skyrmion-hosting chiral magnets include the high-$T_c$, $\beta$-Mn-type Co-Zn-Mn \cite{tokunaga_new_2015} alloys ($P4_{1}32$) and the multiferroic Cu$_2$OSeO$_3$ \cite{seki_observation_2012} ($P2_{1}3$) -- the target material of the present study.

The extensive study of skyrmions has in part been motivated by the observation that they can be moved with low current densities \cite{iwasaki_current-induced_2013}, giving skyrmions prospective applications in future high-density, ultra-low power memory devices. In such devices, skyrmions are required to exist at room temperature and zero applied magnetic field, ensuring the preservation of any stored data while the device is unpowered. The limited extent of the equilibrium SkL phase in temperature and applied field therefore presents a significant obstacle which must be overcome. Metastable skyrmions, which have now been observed in the Co-Zn-Mn alloys \cite{karube_robust_2016, karube_skyrmion_2017, morikawa_deformation_2017}, MnSi \cite{oike_interplay_2016, nakajima_skyrmion_2017, kagawa_current-induced_2017}, Fe$_{0.5}$Co$_{0.5}$Si \cite{milde_unwinding_2013, munzer_skyrmion_2010} and Cu$_2$OSeO$_3$ \cite{okamura_transition_2016, chacon_observation_2018}, offer distinct advantages as a candidate for device application. They are observed to exist over a greatly extended field and temperature range, including at zero field and room temperature in Co$_9$Zn$_9$Mn$_2$ \cite{karube_skyrmion_2017}. Moreover, the reading of metastable skyrmions, and writing using rapid temperature control via current pulses, has already been demonstrated \cite{oike_interplay_2016}. However, their utility depends upon two crucial factors: the metastable SkL state must have a lifetime long enough to prevent data degradation, and the writing process must reliably produce skyrmions at feasible cooling rates.

Previous work has suggested that metastable skyrmion strings unwind by the proliferation and subsequent motion of magnetic Bloch points, which form where skyrmion strings either merge together, or break in two, as illustrated in Fig. \ref{fig1}(a) (1. and 2. respectively) \cite{milde_unwinding_2013}. These Bloch points act as sources or sinks of topological charge, allowing the skyrmion tubes to unwind into the competing conical or helical states \cite{wild_entropy-limited_2017}. It has been speculated that a key factor in the lifetime of the metastable state is that these Bloch points are strongly pinned by defects and disorder present in the underlying crystal lattice \cite{karube_robust_2016}, propagating only by thermally-assisted creep motion \cite{kagawa_current-induced_2017}, in a manner similar to the propagation of magnetic domain walls \cite{parkin_magnetic_2008,kolton_creep_2009,chauve_creep_2000,kleemann_modes_2007}. This effect offers a potential route to engineer the lifetime of metastable skyrmions via doping. In the present study, we quantify the role that the controlled introduction of non-magnetic zinc ions plays on the lifetime of metastable skyrmions in Cu$_2$OSeO$_3$. Our detailed time-resolved AC susceptibility measurements reveal that this both dramatically increases the lifetime and substantially reduces the cooling rate required for the formation of the metastable SkL state. 

\section{Methods}

Polycrystalline Zn-doped Cu$_2$OSeO$_3$ powders were synthesised by solid state reaction, and single crystals were then grown using the chemical vapour transport technique, as detailed in previous work \cite{stefancic_origin_2018}. The elemental composition composition of the crystals was measured by Inductively Coupled Plasma Mass Spectroscopy (ICP-MS). Known masses ($\sim$1-5~mg) of each crystal were dissolved in high purity nitric acid (Romil-SpA) and diluted with high purity water. The resulting solutions were then analysed using an X-Series 2 ICP-MS (Thermo Scientific), considering all naturally occurring isotopes of Cu, Zn and Se to monitor the possible presence of impurity elements and polyatomic species. These measurements revealed the actual Zn substitution levels of the crystals selected for this study to be 2.8\%, 2.5\%, 1.0 \% and 0.0\%. 

To identify the presence of metastable skyrmions, three contrasting field-temperature procedures were employed: zero field-cooling (ZFC), field-cooling (FC), and field-cooling around the equilibrium SkL state (FC*), as illustrated in Fig. \ref{fig1}(b). In the ZFC procedure, the sample was cooled from 70~K down to the final temperature in zero applied magnetic field (solid black arrow). In the FC procedure, the sample was cooled in a field of 22 mT, from 70~K, through the equilibrium SkL state, down to the final temperature (solid green arrow). In the FC* procedure, the sample was cooled at 0 mT from 70~K to 50~K. A field of 22~mT was then applied, and the sample was cooled down to the final temperature (solid red arrows). After initialising the magnetic configuration of the sample with these procedures, measurements were performed with isothermal field-sweeps (dashed black, green and red arrows). The samples were aligned with the magnetic field applied along the [111] direction for all measurements.

Magnetometry measurements were recorded using a superconducting quantum interference device magnetometer (MPMS3, Quantum Design), furnished with the optional AC magnetometry option. All AC susceptibility measurements were performed at a frequency of 10 Hz, and an amplitude of 0.1 mT. Samples were mounted on a quartz rod with GE varnish, and positioned inside the sample chamber. The $T_{\rm c}$ of each sample was measured with magnetisation versus temperature measurements, and were found to be 55.4(1), 56.2(1), 57.4(1) and 58.8(1)~K respectively. Magnetisation versus field measurements determined that the value of the saturation magnetisation was reduced with the addition of non magnetic zinc ions, confirming that the Zn ions were substituted on to the Cu lattice sites. 

The broadband microwave absorption spectroscopy was carried out by using a vector network analyser, with the sample mounted on a coplanar waveguide as described in previous work \cite{okamura_microwave_2013}. At each temperature, the spectrum of microwave absorption caused by magnetic resonance, $\Delta S_{12} (\nu) = S_{12} (\nu) - S^{\rm ref}_{12}(\nu)$, was derived by subtraction of the common background $S^{\rm ref}_{12}(\nu)$ from the raw transmittance spectrum $S_{12} (\nu)$. Here, a spectrum measured at an applied field of 250~mT was adopted as the reference spectrum $S^{\rm ref}_{12}(\nu)$ -- at this field, the magnetic resonance is absent within our target frequency range from 1~GHz to 6~GHz.

\begin{figure*}
\centering
\includegraphics[width=\textwidth]{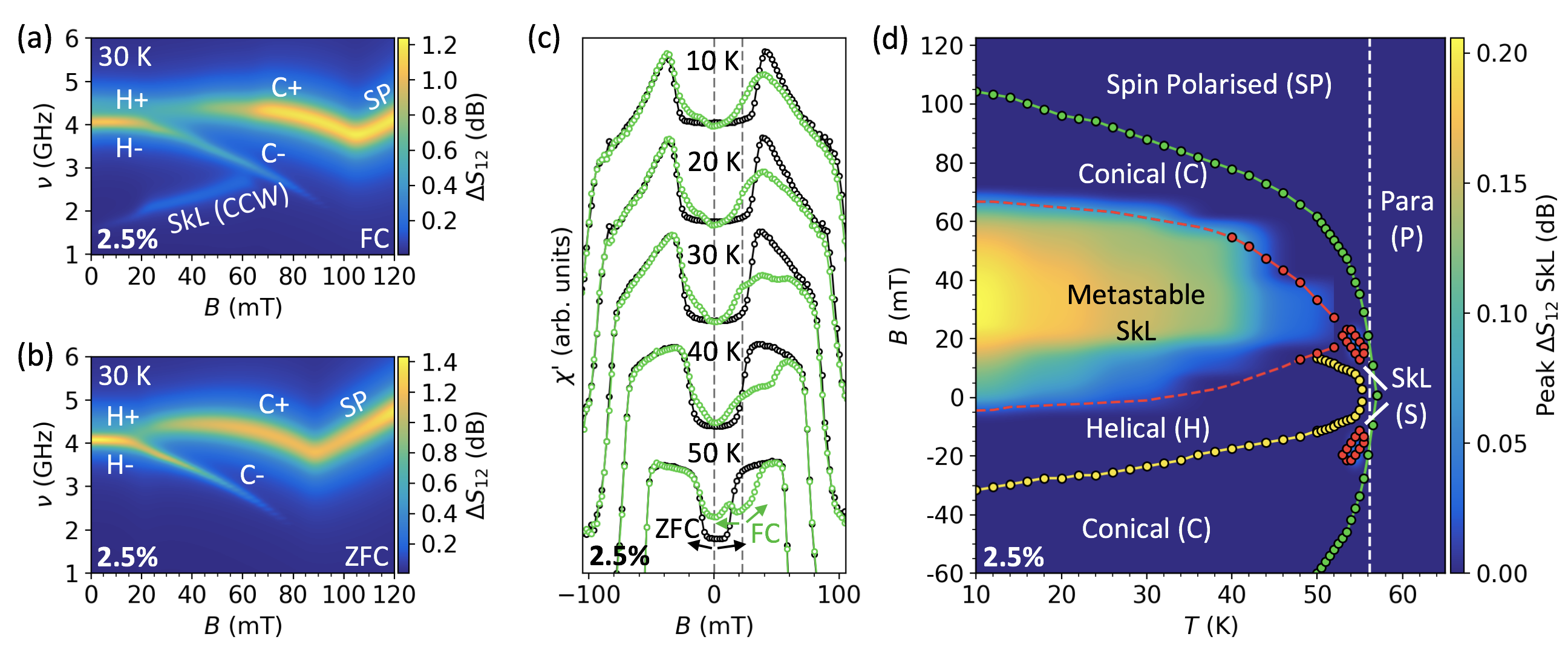}
\caption{(a, b) Microwave absorption spectra measured in the 2.5\% sample as a function of applied magnetic field after the FC and ZFC processes respectively. Magnetic resonances associated with the helical (H$\pm$), conical (C$\pm$) and spin polarised (SP) spin structures are labelled. In (a), the counter-clockwise (CCW) skyrmion lattice (SkL) mode is observed. (c) AC susceptibility measurements performed on the 2.5\% sample after FC (green) and ZFC (black) processes. (d) The magnetic phase diagram of the 2.5\% Zn-doped sample measured after FC measurements. The colourmap shows the strength of the SkL CCW mode measured in the microwave absorption measurements at each temperature and field. The coloured dots show the SkL (red), helical-conical (yellow) and conical-spin polarised (green) phase boundaries, as determined by AC susceptibility measurements.}
\label{fig2}
\end{figure*}

The small angle neutron scattering (SANS) measurements were performed on the D33 instrument at the Institut Laue-Langevin (ILL). The Cu$_2$OSeO$_3$ single crystal was mounted on a 200~$\mu$m thick aluminium plate, positioned inside a 'Blue Charlie' superconducting magnet helium flow cryostat, and aligned with the neutron beam. Measurements were performed with a neutron wavelength of 10~\r{A} and a collimation distance of 12~m. SANS patterns shown are the sum of measurements performed in a rocking curve, achieved by rotating the angle of the field and sample relative to the neutron beam by $\pm$5 degrees and measuring for one minute every 0.5 degrees.

\section{Identifying Metastable Skyrmions}

Microwave absorption spectra, measured as a function of applied field after FC and ZFC to 30~K, are displayed in Figs. 2(a) and 2(b) respectively. In this technique, microwaves are absorbed at the characteristic resonant frequencies of the magnetic structures present in the sample \cite{okamura_microwave_2013,garst_collective_2017, mochizuki_dynamical_2015}. Absorption peaks belonging to the helical (H$\pm$), conical (C$\pm$) and spin polarised (SP) magnetic structures are observed in the response of the sample after both FC and ZFC procedures. However, there is an additional absorption mode with positive $B$-slope in the FC data which we ascribe to the counter-clockwise (CCW) resonance of the metastable skyrmion lattice \cite{onose_observation_2012,okamura_transition_2016}. AC susceptibility, $\chi$', measured at a range of temperatures for both FC and ZFC procedures is shown in Fig. \ref{fig2}(c). There is a clear depression in the measured value of $\chi$' around 22 mT after the FC process, which is consistent with the $\chi$' signal observed for the equilibrium SkL state \cite{bannenberg_magnetic_2016}. Therefore, we attribute this difference to the presence of the metastable SkL state. Results from these two techniques are summarised in the magnetic phase diagram in Fig. \ref{fig2}(d), which demonstrates the large extent of the metastable SkL phase, compared to the smaller equilibrium phase (S) (see Appendix). Here, the colour map plots the peak absorption of the CCW SkL resonance measured at each field and temperature, whilst the coloured dots indicate the boundaries of the magnetic phases as determined by AC susceptibility measurements.

\begin{figure}
\centering
\includegraphics[width=0.45\textwidth]{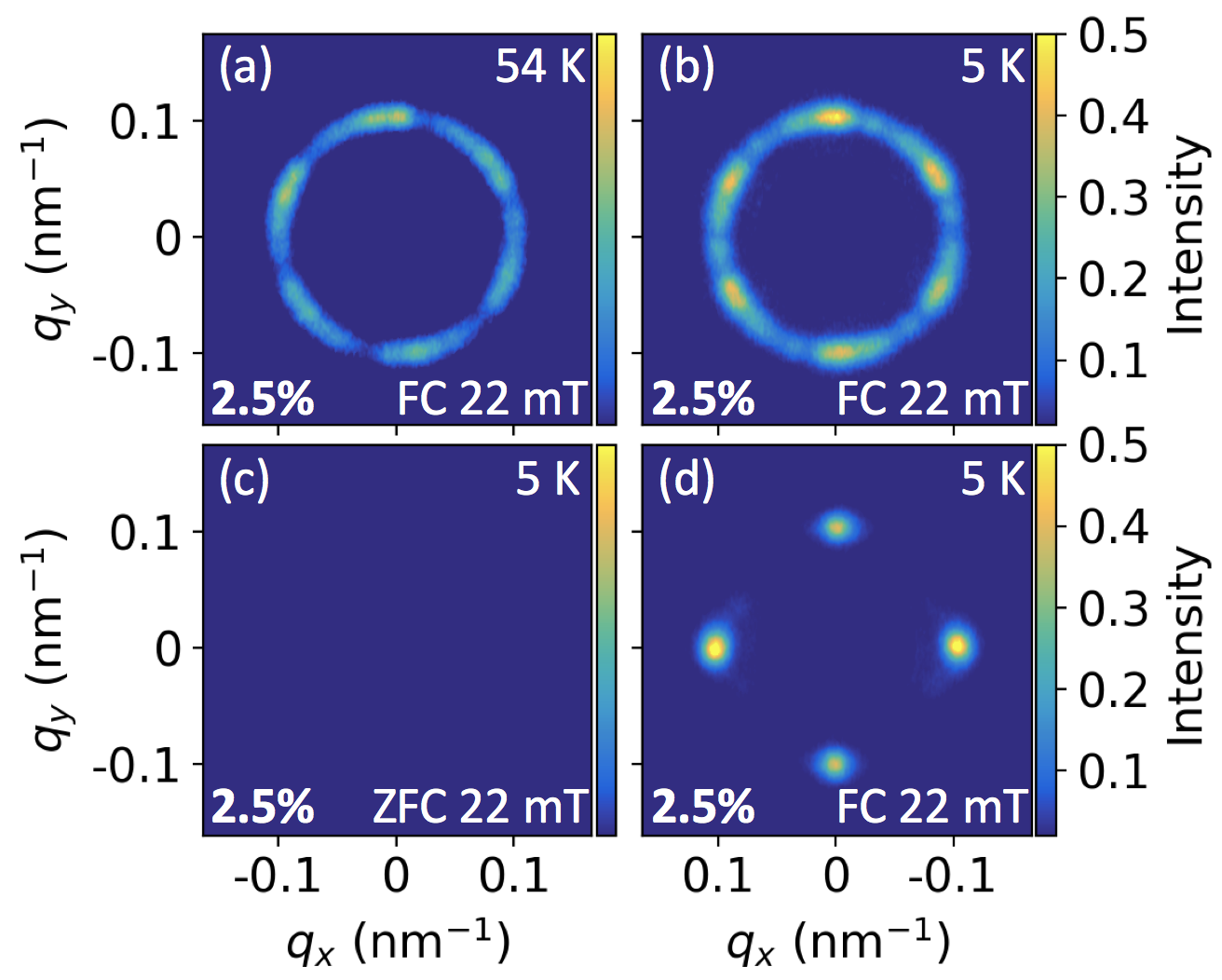}
\caption{(a-d) Small angle neutron scattering (SANS) diffraction patterns recorded at selected temperatures, after both field cooling (a,b,d) and zero field cooling (c) procedures, with the applied magnetic field parallel (a,b,c) and perpendicular (d) to the incident neutron beam.}
\label{fig3}
\end{figure}

Small angle neutron scattering (SANS) measurements were employed to confirm our identification of the metastable SkL state. These measurements were performed on a larger 2.8\% Zn-doped sample. With the magnetic field applied parallel to the neutron beam, Fig. \ref{fig3}(a) displays the sixfold-symmetric pattern characteristic of the hexagonal equilibrium SkL at 54~K, with a skyrmion spacing of $62.6\pm0.1$~nm. Fig. \ref{fig3}(b) shows the SANS pattern recorded after FC down to 5~K. The resulting sixfold pattern confirms the survival of the metastable SkL state at low temperatures, with a spacing of $63.0\pm0.1$~nm. Fig. \ref{fig3}(c) demonstrates the absence of a sixfold pattern at 22~mT after the ZFC procedure to 5~K, indicating no skyrmions are present, as expected. Here, the magnetic state is expected to consist of helices aligned to the [100] crystalline axes. However, as no [100] directions are perpendicular to the neutron beam, we observe no diffraction intensity. Finally, Fig. \ref{fig3}(d) displays the SANS pattern measured after FC to 5 K with the magnetic field applied perpendicular to the neutron beam. Here, the vertical peaks belong to the metastable skyrmion lattice, while the horizontal peaks indicate the coexistence of the competing conical state.

\section{Lifetime of Metastable Skyrmions}

\begin{figure*}
\centering
\includegraphics[width=0.95\textwidth]{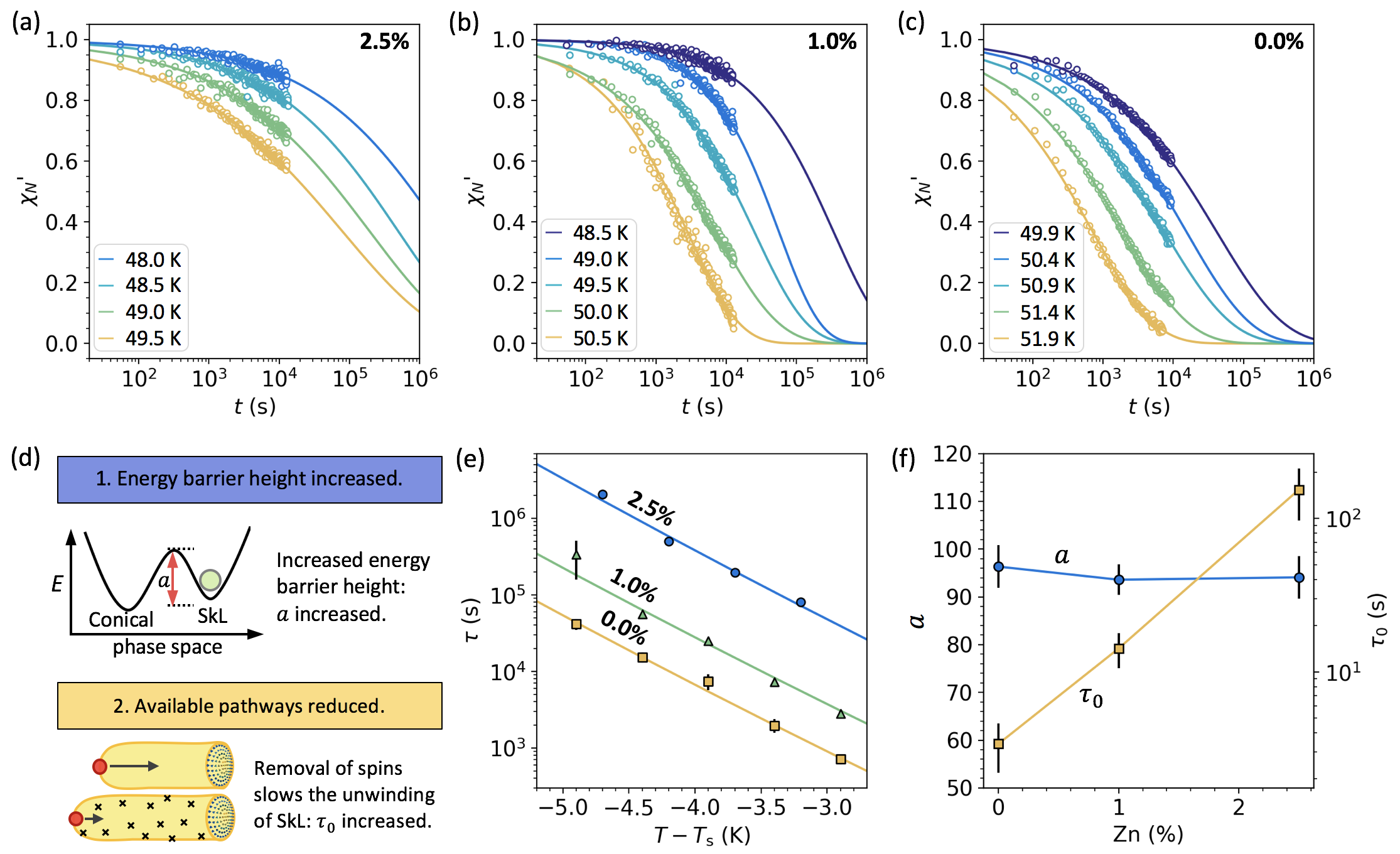}
\caption{(a-c) Normalised, time-resolved AC susceptibility measurements at a range of temperatures for the 2.5\%, 1.0\% and 0.0\% crystals respectively. The datasets are fitted with stretched exponential functions. (d) Schematic representations of the two possible scenarios for increasing the lifetime of metastable skyrmions, based on arguments from Arrhenius’ Law. (e) The extracted lifetimes from (a-c) plotted as a function of $T-T_{\mathrm{s}}$, and fitted with Arrhenius’ Law, determining values for $\tau_0$ and $a$. (f) The fitted $a$ and $\tau_0$ parameters are plotted as a function of the Zn-doping level.}
\label{fig4}
\end{figure*}

Once formed, the population of metastable skyrmions $S(t)$ is observed to follow a decay modelled by a stretched exponential, with a temperature dependent lifetime, $\tau(T)$. During this decay process, the real part of the AC susceptibility, $\chi'$, relaxes from its value in the skyrmion state to the value in the conical state (see Appendix). Assuming that changes in the value of $\chi'$ are proportional to changes in the skyrmion population, the time dependence of the normalised AC susceptibility, $\chi_N'$ (see Fig. \ref{fig8}(b)), can be modelled with a stretched exponential function,
\begin{equation} \label{eq1}
\chi_N'(t) = -\textrm{exp}\left[-\left(\frac{t}{\tau(T)}\right)^\beta\right].
\end{equation}
Utilising this expression and time-resolved measurements of $\chi_N'$, as shown in Fig. \ref{fig4}(a-c), the lifetime of the metastable state was measured as a function of temperature for each Zn-doped crystal. Fig. \ref{fig4}(e) displays the measured values of $\tau(T)$ for each crystal, plotted as a function of $T-T_{\rm{s}}$, where $T_{\rm{s}}$ is the lowest temperature extent of the equilibrium skyrmion state, taken as 4~K below $T_{\rm{c}}$ for all crystals. It is evident that for a given $T-T_{\rm{s}}$, the 2.5\% crystal has a lifetime longer by a factor of ~50 when compared to the 0\% crystal. 

\begin{figure*}
\centering
\includegraphics[width=0.95\linewidth]{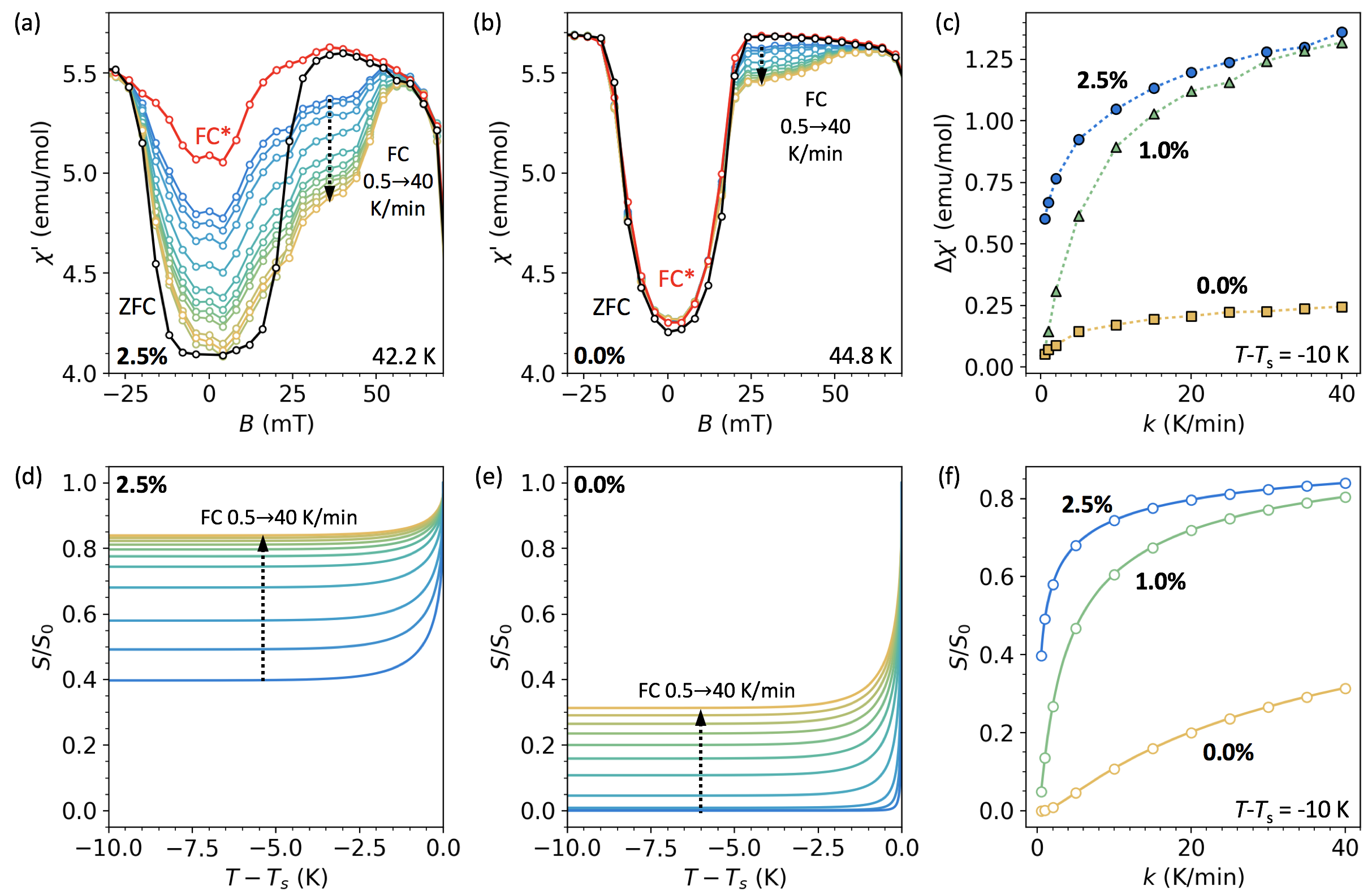}
\caption{(a, b) AC susceptibility, $\chi'$, plotted as a function of applied magnetic field for the 2.5\% and 0.0\% Zn-doped samples respectively, measured after ZFC (black) and FC* (red) procedures at 10 K below $T_{\mathrm{s}}$. Multiple FC datasets are also shown, measured after different cooling rates between 0.5 (blue) and 40 (yellow) K/min. (c) The difference between the FC and FC* $\chi'$ data, $\Delta \chi'$, measured at 22 mT in all samples, plotted as a function of cooling rate. (d, e) The simulated evolution of the metastable skyrmion population for the 2.5\% and 0.0\% crystals, plotted as a function of temperature below $T_{\mathrm{s}}$ at different cooling rates, calculated using values for $a$ and $\tau_0$ from Fig. \ref{fig3}. (f) The final simulated population of metastable skyrmions at 10 K below $T_{\mathrm{s}}$, plotted as a function of cooling rate for all three samples. The circles show the simulated values at the same cooling rates measured in (c).}
\label{fig5}
\end{figure*}

The relationship between the lifetime of the metastable SkL state and the temperature of the sample allows us to model the measured lifetimes using Arrhenius' Law \cite{karube_skyrmion_2017},
\begin{equation} \label{eq2}
\tau(T)=\tau_0\textrm{exp}\left[\frac{-E_b}{k_{\rm{B}}T}\right]=\tau_0\textrm{exp}\left[a\frac{(T_{\mathrm{s}}-T)}{T}\right].
\end{equation}
Here, the typical activation energy term has been replaced with the energy barrier $E_b$ protecting the metastable SkL. As the sample approaches $T_{\rm{s}}$, the height of $E_b$, and therefore the lifetime, are reduced. For the SkL to conical transition, the temperature dependence of the energy barrier can be approximated as $E_b/k_{\rm{B}} = a(T-T_{\rm{s}})$, with $a$ as the linear proportional constant \cite{oike_interplay_2016}. Using this framework, and the model of skrymion strings unwinding via Bloch point formation and motion, we present two possible ways in which the lifetime of the metastable SkL state can be increased with doping, as depicted in Fig. \ref{fig4}(d). In Scenario 1, the value of $a$ is increased, corresponding to an increase in the energy barrier which must be overcome to unwind the skyrmion state. In Scenario 2, the value of $\tau_0$ is increased. An important component of this prefactor, is an entropic correction, which concerns the available pathways across an activation barrier \cite{wild_entropy-limited_2017}. Therefore, if there are fewer available pathways, the lifetime of the metastable state would be increased.

We utilise equation \ref{eq2} to fit the datasets in Fig. \ref{fig4}(e) and extract the gradient, $a$, and the intercept, $\tau_0$, for each crystal. Fig. \ref{fig4}{f} displays these fitted parameters, plotted as a function of Zn-doping level. It is clear that $a$ is constant across all the crystals within experimental uncertainty, with an average value of $95\pm3$ which corresponds to an $E_b/k_{\rm{B}}$ of $\sim5\times10^3$ K at 0~K. The energy barrier height is therefore not substantially altered by the introduction of zinc ions. In contrast, $\tau_0$, plotted on the secondary, logarithmic axis of Fig. \ref{fig3}(f), exhibits a non-linear increase with Zn-doping, with values of $150\pm50$, $14\pm4$ and $3\pm1$ seconds for the 2.5\%, 1.0\% and 0.0\% crystals respectively. This suggests that the removal of spin sites due to the inclusion of non-magnetic Zn ions may limit the number of available pathways by which the Bloch points can overcome the energy barrier to unwind the skyrmion strings. The number of possible spin configurations in a magnetic system typically scales with the number of spins as a power law, and we may expect that the number of pathways between two configurations to also scale as a power law. The dramatic, non-linear, increase of $\tau_0$ with doping could therefore be explained by the expected reduction in the number of available pathways as one removes spins from the system. 

The fraction of metastable skyrmions which survives after FC, $S_0$, is related to the cooling rate, $k$. Because the lifetime of the metastable SkL state is shortest just below $T_{\rm{s}}$, cooling through this region slowly results in a considerable loss of skyrmion population. We investigated this relationship by measuring $\chi'$ after FC at a range of $k$ between 0.5 and 40 K/min. The results for the 2.5\% and 0.0\% crystals, measured 10~K below $T_{\rm{s}}$, are plotted in Fig. \ref{fig5}(a,b) respectively (blue-yellow points). These measurements are contrasted with data measured after the ZFC (black) and FC* (red) procedures. We assume that the difference in $\chi'$ between the FC and FC* processes, $\Delta\chi'=\chi_{\rm{{FC}^*}}'-\chi_{\mathrm{FC}}'$, is proportional to the population of metastable skyrmions. In Fig 5(c), the measured value of $\Delta\chi'$ at 22 mT is plotted as a function of cooling rate for all three crystals. For all cooling rates, the value of $\Delta\chi'$ in the 2.5\% crystal is far greater than that of the 0.0\% crystal, and we therefore infer that the population of metastable skyrmions is substantially higher.

In order to estimate the metastable skyrmion population in each sample from this experimental data, we calculated the expected metastable skyrmion population loss during FC. Our derivation (see Appendix) yields an expression for the metastable population after cooling from $T_{\rm{s}}$ to the final temperature $T_f$ at a rate of $k$,
\begin{equation} \label{eq3}
S_0=S_i \textrm{exp}\left[ \int\limits_{T_{\mathrm{s}}}^{T_f}\frac{-\beta}{T-T_{\mathrm{s}}}\left(\frac{T-T_{\mathrm{s}}}{k\tau_0\textrm{exp}\left[a\frac{T_{\mathrm{s}}-T}{T}\right]}\right)^\beta dT\right],
\end{equation}
where $S_i$ is the initial population. We use this model, and the values for $\beta$, $a$ and $\tau_0$ fitted for each crystal in Fig. \ref{fig4}, to simulate the evolution of the skyrmion population as a function of temperature for cooling rates between 0.5 and 40 K/min. The results for the 2.5\% and 0.0\% crystals are displayed in Fig. \ref{fig5}(d,e). It is clear that largest population loss occurs in the first $\sim$2~K below $T_{\rm{s}}$, after which the long lifetime effectively locks in the population. The final calculated metastable skyrmion population $S_0/S_i$ for each crystal at $(T-T_{\rm{s}}= 10 K$ is plotted as a function of $k$ in Fig. \ref{fig5}(f). The strong qualitative agreement between the experimental data in Fig. \ref{fig5}(c) and the simulated model in Fig. \ref{fig5}(f) suggests that the Arrhenius model for $\tau(T)$ is valid, and that $E_b/k_{\rm{B}} = a(T_{\mathrm{s}}-T)$ is a reasonable assumption for the relationship between energy barrier and temperature. These graphs show that although a cooling rate as high as 40~K/min cannot achieve a metastable population of $\sim$50\% in the 0.0\% crystal, this is achieved at cooling rate of just 1~K/min in the 2.5\% crystal.

We have shown that doping Zn ions onto the Cu sites in Cu$_2$OSeO$_3$ crystals increases the lifetime of the metastable SkL state. As a result, the cooling rate required to achieve a substantial metastable population when field-cooling is greatly reduced. Our analysis of lifetimes measured as a function of temperature suggests that the removal of spins caused by the substitution of the magnetic Cu ions with non-magnetic Zn ions is responsible for this increased lifetime, limiting the number of available pathways by which the metastable skyrmion strings can unwind. We expect that this effect can be exploited to engineer the metastable SkL lifetime in other doped skyrmion-hosting systems.

\begin{acknowledgments}
The authors would like to thank K. Frank\'{e} and T. Hicken for valuable discussions about the work. Appreciation is also given to M. Sussmuth at the I10 beamline, Diamond Light Source. This work was supported by the UK Skyrmion Project EPSRC Programme Grant (EP/N032128/1), ILL proposal 5-41-982, and the JSPS Summer Program 2017 in collaboration with SOKENDAI, who funded M. T. Birch's placement at RIKEN, Japan. S. Seki was supported by Grants-In-Aid for Scientific Research (A) (Grant No. 18H03685) and Grant-in-Aid for Scientific Research on Innovative Area, "Nano Spin Conversion Science" (Grant No.17H05186) from JSPS, and PRESTO from JST. M. N. Wilson acknowledges the support of the Natural Sciences and Engineering Research Council of Canada (NSERC).\\
\end{acknowledgments}

\appendix*
\section*{\textbf{APPENDIX}}
\begin{figure}
\centering
\includegraphics[width=0.45\textwidth]{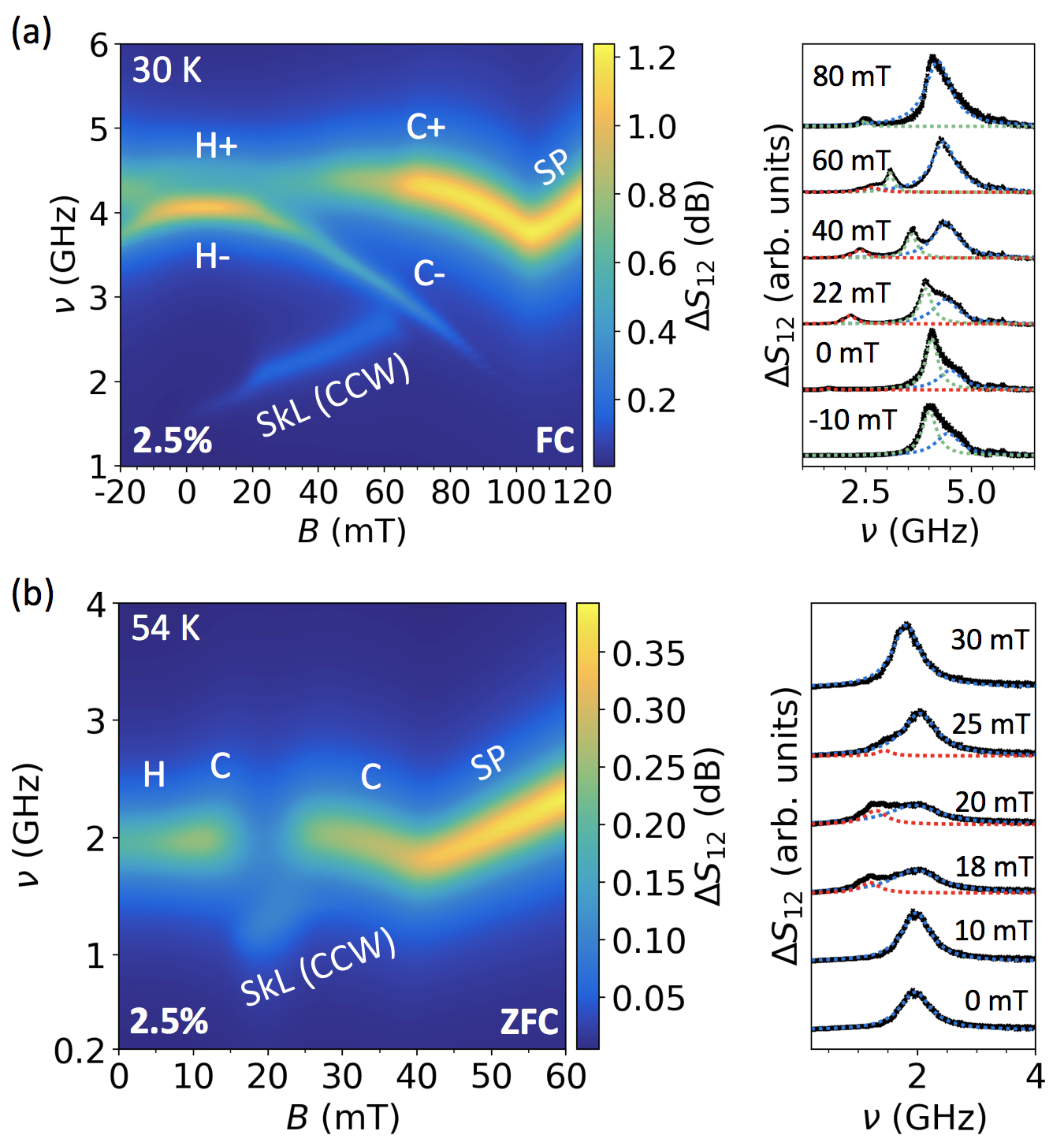}
\caption{(a) Microwave absorption spectra measured on the 2.5\% sample as a function of field at 30~K after FC (left panel). Individual spectra measured at each field, with peaks belonging to the metastable SkL (red) H/C- (green) and H/C+ (blue) resonances fitted with Lorentzian functions (right). (b) Microwave spectra recorded as a function of field after ZFC to 54~K (left). The CCW SkL peak here is the equilibrium SkL phase. Peaks belonging to the SkL (red) H/C (blue) resonances are fitted with Lorentzian functions (right).
}
\label{fig7}
\end{figure}
 
Figs. \ref{fig7}(a) displays one of the microwave absorption spectra used to plot the phase diagram in Fig. \ref{fig2}(d) for the 2.5\% sample. After FC at 22~mT to 30~K, absorption spectra were measured between -20 and 120 mT, $\Delta S_{12}(\nu)$. After the background subtraction, each spectrum was fitted using multiple Lorentzian peaks (coloured lines). This is demonstrated in the spectra plotted alongside each colourmap, which display the data (black) along with the fitted Lorentzian peaks. Fig. \ref{fig7}(b) displays the microwave spectra when measuring a field sweep at 54 K, passing through the equilibrium SkL phase, again fitted with Lorentzian peaks. At this temperature, the H$\pm$ and C$\pm$ resonances are indistinguishable. The fitted peak height of the CCW SkL resonance at each temperature and field was used to plot the colourmap in Fig. \ref{fig2}(d).

\begin{figure}
\centering
\includegraphics[width=0.35\textwidth]{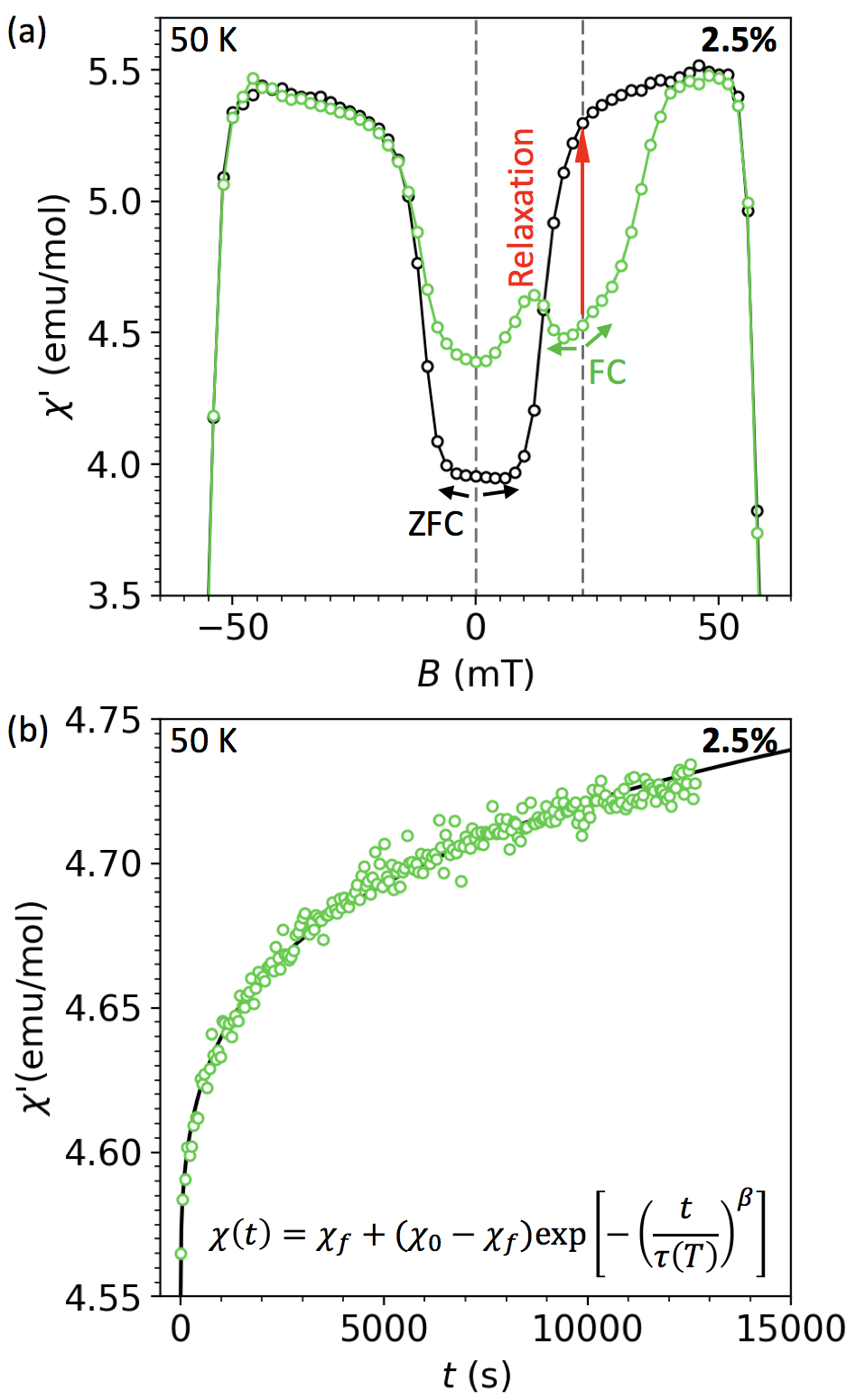}
\caption{(a) AC susceptibility, $\chi'$ measured as a function of applied magnetic field after ZFC (black) and FC (green) at 50 K in the 2.5\% crystal. The expected relaxation of $\chi'$ the metastable SkL decays is illustrated by the red arrow. (b) The value of $\chi'(t)$ measured at 50 K in the 2.5\% sample is plotted as a function of time. The data was fitted using the stretched exponential shown.}
\label{fig8}
\end{figure}

Fig. \ref{fig8}(a) displays AC susceptibility data measured at 50 K in the 2.5\% sample after both FC (green) and ZFC (black). The red arrow indicates how the AC susceptibility $\chi'(t)$ relaxes as the metastable SkL state decays into the conical state with time. The time evolution of the metastable skyrmion population follows a stretched exponential function, 
\begin{equation}
S(t)=S_0 \textrm{exp}\left[-\left(\frac{t}{\tau(T)}\right)^\beta\right].
\label{eqA1}
\tag{A1}
\end{equation}
By assuming that changes in $\chi'(t)$ are proportional to changes in S(t), we can model the evolution of the AC susceptibility data with $\chi'(t)=\chi_f+(\chi_0-\chi_f)\textrm{exp}\left[-\left(t/\tau(T)\right)\right]$, where $\chi_0$ is the initial value of $\chi'$ at $t=0$, and $\chi_f$ is the value the system is tending to at $t=\infty$. In the main text, this is simplified by normalising the AC susceptibility, as shown in equation \ref{eq1}. The raw data measured in one lifetime measurement is displayed in Fig. \ref{fig8}(b). Using the above equation, the data has been fitted to extract values of $\tau$ and $\beta$. The parameter $\beta$ gives an indication of the spread of lifetimes in the sample, and was found to be between 0.4 and 0.6, depending upon the sample and temperature.

To simulate the loss of metastable skyrmion population as the sample is cooled, we derived the following model. We begin with the expression for a population of metastable skyrmions, $S$, decaying over time $t$ with a temperature dependent lifetime $\tau(T)$, as shown in equation (4). Unlike a standard exponential function which is self-similar, when  $\beta\neq 1$ the shape of a stretched exponential function changes with time. To account for this, we consider the population decay during the cooling process as a series of time steps with duration $\Delta t$, each starting at a time $t_n = n \Delta t$. For a single time step, this gives,
\begin{equation} \label{eqA2}
\frac{S_{n+1}}{S_{n}}=\frac{\textrm{exp}\left[-\left(\frac{t_n+\Delta t}{\tau(T)}\right)^\beta\right]}{\textrm{exp}\left[-\left(\frac{t_n}{\tau(T)}\right)^\beta\right]}.
\tag{A2}
\end{equation}
It follows that the final population, $S_f$, after $N$ time steps is a product series,
\begin{equation} \label{eqA3}
\displaystyle S_f = S_i \prod_{n=1}^{N} \textrm{exp}\left[\left(\frac{t_n}{\tau(T)}\right)^\beta-\left(\frac{t_n+\Delta t}{\tau(T)}\right)^\beta\right],
\tag{A3}
\end{equation}
\begin{equation} \label{eqA4}
S_f = S_i \textrm{exp}\left[\sum_{n=1}^{N}\left(\frac{t_n}{\tau(T)}\right)^\beta-\left(\frac{t_n+\Delta t}{\tau(T)}\right)^\beta\right].
\tag{A4}
\end{equation}
We now define the sum inside of the exponential function to be $g(t)$, and take the limit as $\Delta t \to 0$,
\begin{equation}\label{eqA5}
\begin{split} 
\lim_{\Delta t\to0} g(t) = & \lim_{\Delta t\to0}  \sum_{n=1}^{N}\left(\frac{t_n}{\tau(T)}\right)^\beta\\ 
 - & \lim_{\Delta t\to0} \sum_{n=1}^{N}   \left(\frac{t_n+\Delta t}{\tau(T)}\right)^\beta
\end{split}
\tag{A5}
\end{equation}

Performing a Taylor expansion of the second term around $\Delta t=0$ yields,
\begin{equation}\label{eqA6}
\begin{split} 
\lim_{\Delta t\to0} g(t) = \lim_{\Delta t\to0} \sum_{n=1}^{N}\left(\frac{t_n}{\tau(T)}\right)^\beta \\
- \lim_{\Delta t\to0} \sum_{n=1}^{N}\left[ \left(\frac{t_n}{\tau(T)}\right)^\beta+\beta\frac{\Delta t}{t_n}\left(\frac{t_n}{\tau(T)}\right)^\beta \right]&+\mathcal{O}^2 .
\end{split}
\tag{A6}
\end{equation}
The higher order terms of $\Delta t^2$ are infinitesimally small as $\Delta t \to 0$ leaving,
\begin{equation} \label{eqA7}
\lim_{\Delta t\to0} g(t) = -\lim_{\Delta t\to0}\sum_{n=1}^{N} \beta\frac{\Delta t}{t_n}\left(\frac{t_n}{\tau(T)}\right)^\beta
\tag{A7}
\end{equation}
\begin{equation} \label{eqA8}
\lim_{\Delta t\to0} g(t) = -\beta \int_{0}^{t_f}\frac{1}{t}\left(\frac{t}{\tau(T)}\right)^\beta dt,
\tag{A8}
\end{equation}
where $t_f$ is the time at the end of the cooling process. Finally, we substitute back into our equation for $S_f$, to find an expression for the population of metastable skyrmions as a function of time,
\begin{equation} \label{eqA9}
S_f=S_i \textrm{exp}\left[-\beta \int_{0}^{t_f}\frac{1}{t}\left(\frac{t}{\tau(T)}\right)^\beta dt\right].
\tag{A9}
\end{equation}

In our analysis, we assume that the lifetime depends upon the temperature of according to our modified Arrhenius' law,
\begin{equation} \label{eqA10}
\tau(T)=\tau_0\textrm{exp}\left[a\left(\frac{T_{\rm s}-T}{T}\right)\right],
\tag{A10}
\end{equation}
where $T_{\rm s}$ is the lowest extent of the equilibrium skyrmiom phase in temperature. With a linear cooling rate, $k<0$, starting at $T=T_{\rm s}$, we can write $T=T_{\rm s}+kt$. We substitute equation (8) into equation (7), and perform a change of variables from $t$ to $T$ to reveal the dependence of $S_f$ on the cooling rate $k$,
\begin{equation} \label{eqA11}
S_0=S_i \textrm{exp}\left[ \int\limits_{T_{\mathrm{s}}}^{T_f}\frac{-\beta}{T-T_{\mathrm{s}}}\left(\frac{T-T_{\mathrm{s}}}{k\tau_0\textrm{exp}\left[a\frac{T_{\mathrm{s}}-T}{T}\right]}\right)^\beta dT\right],
\tag{A11}
\end{equation}
\vspace{\baselineskip}\linebreak

\bibliography{sample}

\begin{thebibliography}{31}%
\makeatletter
\providecommand \@ifxundefined [1]{%
 \@ifx{#1\undefined}
}%
\providecommand \@ifnum [1]{%
 \ifnum #1\expandafter \@firstoftwo
 \else \expandafter \@secondoftwo
 \fi
}%
\providecommand \@ifx [1]{%
 \ifx #1\expandafter \@firstoftwo
 \else \expandafter \@secondoftwo
 \fi
}%
\providecommand \natexlab [1]{#1}%
\providecommand \enquote  [1]{``#1''}%
\providecommand \bibnamefont  [1]{#1}%
\providecommand \bibfnamefont [1]{#1}%
\providecommand \citenamefont [1]{#1}%
\providecommand \href@noop [0]{\@secondoftwo}%
\providecommand \href [0]{\begingroup \@sanitize@url \@href}%
\providecommand \@href[1]{\@@startlink{#1}\@@href}%
\providecommand \@@href[1]{\endgroup#1\@@endlink}%
\providecommand \@sanitize@url [0]{\catcode `\\12\catcode `\$12\catcode
  `\&12\catcode `\#12\catcode `\^12\catcode `\_12\catcode `\%12\relax}%
\providecommand \@@startlink[1]{}%
\providecommand \@@endlink[0]{}%
\providecommand \url  [0]{\begingroup\@sanitize@url \@url }%
\providecommand \@url [1]{\endgroup\@href {#1}{\urlprefix }}%
\providecommand \urlprefix  [0]{URL }%
\providecommand \Eprint [0]{\href }%
\providecommand \doibase [0]{http://dx.doi.org/}%
\providecommand \selectlanguage [0]{\@gobble}%
\providecommand \bibinfo  [0]{\@secondoftwo}%
\providecommand \bibfield  [0]{\@secondoftwo}%
\providecommand \translation [1]{[#1]}%
\providecommand \BibitemOpen [0]{}%
\providecommand \bibitemStop [0]{}%
\providecommand \bibitemNoStop [0]{.\EOS\space}%
\providecommand \EOS [0]{\spacefactor3000\relax}%
\providecommand \BibitemShut  [1]{\csname bibitem#1\endcsname}%
\let\auto@bib@innerbib\@empty
\bibitem [{\citenamefont {Nagaosa}\ and\ \citenamefont
  {Tokura}(2013)}]{nagaosa_topological_2013}%
  \BibitemOpen
  \bibfield  {author} {\bibinfo {author} {\bibfnamefont {N.}~\bibnamefont
  {Nagaosa}}\ and\ \bibinfo {author} {\bibfnamefont {Y.}~\bibnamefont
  {Tokura}},\ }\href {\doibase 10.1038/nnano.2013.243} {\bibfield  {journal}
  {\bibinfo  {journal} {Nature Nanotechnology}\ }\textbf {\bibinfo {volume}
  {8}},\ \bibinfo {pages} {899} (\bibinfo {year} {2013})}\BibitemShut {NoStop}%
\bibitem [{\citenamefont {M{\"u}hlbauer}\ \emph {et~al.}(2009)\citenamefont
  {M{\"u}hlbauer}, \citenamefont {Binz}, \citenamefont {Jonietz}, \citenamefont
  {Pfleiderer}, \citenamefont {Rosch}, \citenamefont {Neubauer}, \citenamefont
  {Georgii},\ and\ \citenamefont {BÃ¶ni}}]{muhlbauer_skyrmion_2009}%
  \BibitemOpen
  \bibfield  {author} {\bibinfo {author} {\bibfnamefont {S.}~\bibnamefont
  {M{\"u}hlbauer}}, \bibinfo {author} {\bibfnamefont {B.}~\bibnamefont {Binz}},
  \bibinfo {author} {\bibfnamefont {F.}~\bibnamefont {Jonietz}}, \bibinfo
  {author} {\bibfnamefont {C.}~\bibnamefont {Pfleiderer}}, \bibinfo {author}
  {\bibfnamefont {A.}~\bibnamefont {Rosch}}, \bibinfo {author} {\bibfnamefont
  {A.}~\bibnamefont {Neubauer}}, \bibinfo {author} {\bibfnamefont
  {R.}~\bibnamefont {Georgii}}, \ and\ \bibinfo {author} {\bibfnamefont
  {P.}~\bibnamefont {BÃ¶ni}},\ }\href {\doibase 10.1126/science.1166767}
  {\bibfield  {journal} {\bibinfo  {journal} {Science}\ }\textbf {\bibinfo
  {volume} {323}},\ \bibinfo {pages} {915} (\bibinfo {year}
  {2009})}\BibitemShut {NoStop}%
\bibitem [{\citenamefont {Yu}\ \emph {et~al.}(2010)\citenamefont {Yu},
  \citenamefont {Onose}, \citenamefont {Kanazawa}, \citenamefont {Park},
  \citenamefont {Han}, \citenamefont {Matsui}, \citenamefont {Nagaosa},\ and\
  \citenamefont {Tokura}}]{yu_real-space_2010}%
  \BibitemOpen
  \bibfield  {author} {\bibinfo {author} {\bibfnamefont {X.~Z.}\ \bibnamefont
  {Yu}}, \bibinfo {author} {\bibfnamefont {Y.}~\bibnamefont {Onose}}, \bibinfo
  {author} {\bibfnamefont {N.}~\bibnamefont {Kanazawa}}, \bibinfo {author}
  {\bibfnamefont {J.~H.}\ \bibnamefont {Park}}, \bibinfo {author}
  {\bibfnamefont {J.~H.}\ \bibnamefont {Han}}, \bibinfo {author} {\bibfnamefont
  {Y.}~\bibnamefont {Matsui}}, \bibinfo {author} {\bibfnamefont
  {N.}~\bibnamefont {Nagaosa}}, \ and\ \bibinfo {author} {\bibfnamefont
  {Y.}~\bibnamefont {Tokura}},\ }\href {\doibase 10.1038/nature09124}
  {\bibfield  {journal} {\bibinfo  {journal} {Nature}\ }\textbf {\bibinfo
  {volume} {465}},\ \bibinfo {pages} {901} (\bibinfo {year}
  {2010})}\BibitemShut {NoStop}%
\bibitem [{\citenamefont {Yu}\ \emph {et~al.}(2011)\citenamefont {Yu},
  \citenamefont {Kanazawa}, \citenamefont {Onose}, \citenamefont {Kimoto},
  \citenamefont {Zhang}, \citenamefont {Ishiwata}, \citenamefont {Matsui},\
  and\ \citenamefont {Tokura}}]{yu_near_2011}%
  \BibitemOpen
  \bibfield  {author} {\bibinfo {author} {\bibfnamefont {X.~Z.}\ \bibnamefont
  {Yu}}, \bibinfo {author} {\bibfnamefont {N.}~\bibnamefont {Kanazawa}},
  \bibinfo {author} {\bibfnamefont {Y.}~\bibnamefont {Onose}}, \bibinfo
  {author} {\bibfnamefont {K.}~\bibnamefont {Kimoto}}, \bibinfo {author}
  {\bibfnamefont {W.~Z.}\ \bibnamefont {Zhang}}, \bibinfo {author}
  {\bibfnamefont {S.}~\bibnamefont {Ishiwata}}, \bibinfo {author}
  {\bibfnamefont {Y.}~\bibnamefont {Matsui}}, \ and\ \bibinfo {author}
  {\bibfnamefont {Y.}~\bibnamefont {Tokura}},\ }\href {\doibase
  10.1038/nmat2916} {\bibfield  {journal} {\bibinfo  {journal} {Nature
  Materials}\ }\textbf {\bibinfo {volume} {10}},\ \bibinfo {pages} {106}
  (\bibinfo {year} {2011})}\BibitemShut {NoStop}%
\bibitem [{\citenamefont {R{\"o}{\ss}ler}\ \emph {et~al.}(2006)\citenamefont
  {R{\"o}{\ss}ler}, \citenamefont {Bogdanov},\ and\ \citenamefont
  {Pfleiderer}}]{rosler_spontaneous_2006}%
  \BibitemOpen
  \bibfield  {author} {\bibinfo {author} {\bibfnamefont {U.~K.}\ \bibnamefont
  {R{\"o}{\ss}ler}}, \bibinfo {author} {\bibfnamefont {A.~N.}\ \bibnamefont
  {Bogdanov}}, \ and\ \bibinfo {author} {\bibfnamefont {C.}~\bibnamefont
  {Pfleiderer}},\ }\href {\doibase 10.1038/nature05056} {\bibfield  {journal}
  {\bibinfo  {journal} {Nature}\ }\textbf {\bibinfo {volume} {442}},\ \bibinfo
  {pages} {797} (\bibinfo {year} {2006})}\BibitemShut {NoStop}%
\bibitem [{\citenamefont {Chacon}\ \emph {et~al.}(2018)\citenamefont {Chacon},
  \citenamefont {Heinen}, \citenamefont {Halder}, \citenamefont {Bauer},
  \citenamefont {Simeth}, \citenamefont {Mühlbauer}, \citenamefont {Berger},
  \citenamefont {Garst}, \citenamefont {Rosch},\ and\ \citenamefont
  {Pfleiderer}}]{chacon_observation_2018}%
  \BibitemOpen
  \bibfield  {author} {\bibinfo {author} {\bibfnamefont {A.}~\bibnamefont
  {Chacon}}, \bibinfo {author} {\bibfnamefont {L.}~\bibnamefont {Heinen}},
  \bibinfo {author} {\bibfnamefont {M.}~\bibnamefont {Halder}}, \bibinfo
  {author} {\bibfnamefont {A.}~\bibnamefont {Bauer}}, \bibinfo {author}
  {\bibfnamefont {W.}~\bibnamefont {Simeth}}, \bibinfo {author} {\bibfnamefont
  {S.}~\bibnamefont {Mühlbauer}}, \bibinfo {author} {\bibfnamefont
  {H.}~\bibnamefont {Berger}}, \bibinfo {author} {\bibfnamefont
  {M.}~\bibnamefont {Garst}}, \bibinfo {author} {\bibfnamefont
  {A.}~\bibnamefont {Rosch}}, \ and\ \bibinfo {author} {\bibfnamefont
  {C.}~\bibnamefont {Pfleiderer}},\ }\href {\doibase 10.1038/s41567-018-0184-y}
  {\bibfield  {journal} {\bibinfo  {journal} {Nature Physics}\ ,\ \bibinfo
  {pages} {1745}} (\bibinfo {year} {2018})}\BibitemShut {NoStop}%
\bibitem [{\citenamefont {Karube}\ \emph {et~al.}(2018)\citenamefont {Karube},
  \citenamefont {White}, \citenamefont {Morikawa}, \citenamefont {Dewhurst},
  \citenamefont {Cubitt}, \citenamefont {Kikkawa}, \citenamefont {Yu},
  \citenamefont {Tokunaga}, \citenamefont {Arima}, \citenamefont {Rønnow},
  \citenamefont {Tokura},\ and\ \citenamefont
  {Taguchi}}]{karube_disordered_2018}%
  \BibitemOpen
  \bibfield  {author} {\bibinfo {author} {\bibfnamefont {K.}~\bibnamefont
  {Karube}}, \bibinfo {author} {\bibfnamefont {J.~S.}\ \bibnamefont {White}},
  \bibinfo {author} {\bibfnamefont {D.}~\bibnamefont {Morikawa}}, \bibinfo
  {author} {\bibfnamefont {C.~D.}\ \bibnamefont {Dewhurst}}, \bibinfo {author}
  {\bibfnamefont {R.}~\bibnamefont {Cubitt}}, \bibinfo {author} {\bibfnamefont
  {A.}~\bibnamefont {Kikkawa}}, \bibinfo {author} {\bibfnamefont
  {X.}~\bibnamefont {Yu}}, \bibinfo {author} {\bibfnamefont {Y.}~\bibnamefont
  {Tokunaga}}, \bibinfo {author} {\bibfnamefont {T.-h.}\ \bibnamefont {Arima}},
  \bibinfo {author} {\bibfnamefont {H.~M.}\ \bibnamefont {Rønnow}}, \bibinfo
  {author} {\bibfnamefont {Y.}~\bibnamefont {Tokura}}, \ and\ \bibinfo {author}
  {\bibfnamefont {Y.}~\bibnamefont {Taguchi}},\ }\href {\doibase
  10.1126/sciadv.aar7043} {\bibfield  {journal} {\bibinfo  {journal} {Science
  Advances}\ }\textbf {\bibinfo {volume} {4}},\ \bibinfo {pages} {eaar7043}
  (\bibinfo {year} {2018})}\BibitemShut {NoStop}%
\bibitem [{\citenamefont {Yokouchi}\ \emph {et~al.}(2018)\citenamefont
  {Yokouchi}, \citenamefont {Hoshino}, \citenamefont {Kanazawa}, \citenamefont
  {Kikkawa}, \citenamefont {Morikawa}, \citenamefont {Shibata}, \citenamefont
  {Arima}, \citenamefont {Taguchi}, \citenamefont {Kagawa}, \citenamefont
  {Nagaosa},\ and\ \citenamefont {Tokura}}]{yokouchi_current-induced_2018}%
  \BibitemOpen
  \bibfield  {author} {\bibinfo {author} {\bibfnamefont {T.}~\bibnamefont
  {Yokouchi}}, \bibinfo {author} {\bibfnamefont {S.}~\bibnamefont {Hoshino}},
  \bibinfo {author} {\bibfnamefont {N.}~\bibnamefont {Kanazawa}}, \bibinfo
  {author} {\bibfnamefont {A.}~\bibnamefont {Kikkawa}}, \bibinfo {author}
  {\bibfnamefont {D.}~\bibnamefont {Morikawa}}, \bibinfo {author}
  {\bibfnamefont {K.}~\bibnamefont {Shibata}}, \bibinfo {author} {\bibfnamefont
  {T.-h.}\ \bibnamefont {Arima}}, \bibinfo {author} {\bibfnamefont
  {Y.}~\bibnamefont {Taguchi}}, \bibinfo {author} {\bibfnamefont
  {F.}~\bibnamefont {Kagawa}}, \bibinfo {author} {\bibfnamefont
  {N.}~\bibnamefont {Nagaosa}}, \ and\ \bibinfo {author} {\bibfnamefont
  {Y.}~\bibnamefont {Tokura}},\ }\href {\doibase 10.1126/sciadv.aat1115}
  {\bibfield  {journal} {\bibinfo  {journal} {Science Advances}\ }\textbf
  {\bibinfo {volume} {4}},\ \bibinfo {pages} {eaat1115} (\bibinfo {year}
  {2018})}\BibitemShut {NoStop}%
\bibitem [{\citenamefont {Tokunaga}\ \emph {et~al.}(2015)\citenamefont
  {Tokunaga}, \citenamefont {Yu}, \citenamefont {White}, \citenamefont
  {RÃžnnow}, \citenamefont {Morikawa}, \citenamefont {Taguchi},\ and\
  \citenamefont {Tokura}}]{tokunaga_new_2015}%
  \BibitemOpen
  \bibfield  {author} {\bibinfo {author} {\bibfnamefont {Y.}~\bibnamefont
  {Tokunaga}}, \bibinfo {author} {\bibfnamefont {X.~Z.}\ \bibnamefont {Yu}},
  \bibinfo {author} {\bibfnamefont {J.~S.}\ \bibnamefont {White}}, \bibinfo
  {author} {\bibfnamefont {H.~M.}\ \bibnamefont {RÃžnnow}}, \bibinfo {author}
  {\bibfnamefont {D.}~\bibnamefont {Morikawa}}, \bibinfo {author}
  {\bibfnamefont {Y.}~\bibnamefont {Taguchi}}, \ and\ \bibinfo {author}
  {\bibfnamefont {Y.}~\bibnamefont {Tokura}},\ }\href {\doibase
  10.1038/ncomms8638} {\bibfield  {journal} {\bibinfo  {journal} {Nature
  Communications}\ }\textbf {\bibinfo {volume} {6}},\ \bibinfo {pages} {7638}
  (\bibinfo {year} {2015})}\BibitemShut {NoStop}%
\bibitem [{\citenamefont {Seki}\ \emph {et~al.}(2012)\citenamefont {Seki},
  \citenamefont {Yu}, \citenamefont {Ishiwata},\ and\ \citenamefont
  {Tokura}}]{seki_observation_2012}%
  \BibitemOpen
  \bibfield  {author} {\bibinfo {author} {\bibfnamefont {S.}~\bibnamefont
  {Seki}}, \bibinfo {author} {\bibfnamefont {X.~Z.}\ \bibnamefont {Yu}},
  \bibinfo {author} {\bibfnamefont {S.}~\bibnamefont {Ishiwata}}, \ and\
  \bibinfo {author} {\bibfnamefont {Y.}~\bibnamefont {Tokura}},\ }\href
  {\doibase 10.1126/science.1214143} {\bibfield  {journal} {\bibinfo  {journal}
  {Science}\ }\textbf {\bibinfo {volume} {336}},\ \bibinfo {pages} {198}
  (\bibinfo {year} {2012})}\BibitemShut {NoStop}%
\bibitem [{\citenamefont {Iwasaki}\ \emph {et~al.}(2013)\citenamefont
  {Iwasaki}, \citenamefont {Mochizuki},\ and\ \citenamefont
  {Nagaosa}}]{iwasaki_current-induced_2013}%
  \BibitemOpen
  \bibfield  {author} {\bibinfo {author} {\bibfnamefont {J.}~\bibnamefont
  {Iwasaki}}, \bibinfo {author} {\bibfnamefont {M.}~\bibnamefont {Mochizuki}},
  \ and\ \bibinfo {author} {\bibfnamefont {N.}~\bibnamefont {Nagaosa}},\ }\href
  {\doibase 10.1038/nnano.2013.176} {\bibfield  {journal} {\bibinfo  {journal}
  {Nature Nanotechnology}\ }\textbf {\bibinfo {volume} {8}},\ \bibinfo {pages}
  {742} (\bibinfo {year} {2013})}\BibitemShut {NoStop}%
\bibitem [{\citenamefont {Karube}\ \emph {et~al.}(2016)\citenamefont {Karube},
  \citenamefont {White}, \citenamefont {Reynolds}, \citenamefont {Gavilano},
  \citenamefont {Oike}, \citenamefont {Kikkawa}, \citenamefont {Kagawa},
  \citenamefont {Tokunaga}, \citenamefont {RÃžnnow}, \citenamefont {Tokura},\
  and\ \citenamefont {Taguchi}}]{karube_robust_2016}%
  \BibitemOpen
  \bibfield  {author} {\bibinfo {author} {\bibfnamefont {K.}~\bibnamefont
  {Karube}}, \bibinfo {author} {\bibfnamefont {J.~S.}\ \bibnamefont {White}},
  \bibinfo {author} {\bibfnamefont {N.}~\bibnamefont {Reynolds}}, \bibinfo
  {author} {\bibfnamefont {J.~L.}\ \bibnamefont {Gavilano}}, \bibinfo {author}
  {\bibfnamefont {H.}~\bibnamefont {Oike}}, \bibinfo {author} {\bibfnamefont
  {A.}~\bibnamefont {Kikkawa}}, \bibinfo {author} {\bibfnamefont
  {F.}~\bibnamefont {Kagawa}}, \bibinfo {author} {\bibfnamefont
  {Y.}~\bibnamefont {Tokunaga}}, \bibinfo {author} {\bibfnamefont {H.~M.}\
  \bibnamefont {RÃžnnow}}, \bibinfo {author} {\bibfnamefont {Y.}~\bibnamefont
  {Tokura}}, \ and\ \bibinfo {author} {\bibfnamefont {Y.}~\bibnamefont
  {Taguchi}},\ }\href {\doibase 10.1038/nmat4752} {\bibfield  {journal}
  {\bibinfo  {journal} {Nature Materials}\ }\textbf {\bibinfo {volume} {15}},\
  \bibinfo {pages} {1237} (\bibinfo {year} {2016})}\BibitemShut {NoStop}%
\bibitem [{\citenamefont {Karube}\ \emph {et~al.}(2017)\citenamefont {Karube},
  \citenamefont {White}, \citenamefont {Morikawa}, \citenamefont {Bartkowiak},
  \citenamefont {Kikkawa}, \citenamefont {Tokunaga}, \citenamefont {Arima},
  \citenamefont {Rønnow}, \citenamefont {Tokura},\ and\ \citenamefont
  {Taguchi}}]{karube_skyrmion_2017}%
  \BibitemOpen
  \bibfield  {author} {\bibinfo {author} {\bibfnamefont {K.}~\bibnamefont
  {Karube}}, \bibinfo {author} {\bibfnamefont {J.~S.}\ \bibnamefont {White}},
  \bibinfo {author} {\bibfnamefont {D.}~\bibnamefont {Morikawa}}, \bibinfo
  {author} {\bibfnamefont {M.}~\bibnamefont {Bartkowiak}}, \bibinfo {author}
  {\bibfnamefont {A.}~\bibnamefont {Kikkawa}}, \bibinfo {author} {\bibfnamefont
  {Y.}~\bibnamefont {Tokunaga}}, \bibinfo {author} {\bibfnamefont
  {T.}~\bibnamefont {Arima}}, \bibinfo {author} {\bibfnamefont {H.~M.}\
  \bibnamefont {Rønnow}}, \bibinfo {author} {\bibfnamefont {Y.}~\bibnamefont
  {Tokura}}, \ and\ \bibinfo {author} {\bibfnamefont {Y.}~\bibnamefont
  {Taguchi}},\ }\href {\doibase 10.1103/PhysRevMaterials.1.074405} {\bibfield
  {journal} {\bibinfo  {journal} {Physical Review Materials}\ }\textbf
  {\bibinfo {volume} {1}},\ \bibinfo {pages} {074405} (\bibinfo {year}
  {2017})}\BibitemShut {NoStop}%
\bibitem [{\citenamefont {Morikawa}\ \emph {et~al.}(2017)\citenamefont
  {Morikawa}, \citenamefont {Yu}, \citenamefont {Karube}, \citenamefont
  {Tokunaga}, \citenamefont {Taguchi}, \citenamefont {Arima},\ and\
  \citenamefont {Tokura}}]{morikawa_deformation_2017}%
  \BibitemOpen
  \bibfield  {author} {\bibinfo {author} {\bibfnamefont {D.}~\bibnamefont
  {Morikawa}}, \bibinfo {author} {\bibfnamefont {X.}~\bibnamefont {Yu}},
  \bibinfo {author} {\bibfnamefont {K.}~\bibnamefont {Karube}}, \bibinfo
  {author} {\bibfnamefont {Y.}~\bibnamefont {Tokunaga}}, \bibinfo {author}
  {\bibfnamefont {Y.}~\bibnamefont {Taguchi}}, \bibinfo {author} {\bibfnamefont
  {T.-h.}\ \bibnamefont {Arima}}, \ and\ \bibinfo {author} {\bibfnamefont
  {Y.}~\bibnamefont {Tokura}},\ }\href {\doibase 10.1021/acs.nanolett.6b04821}
  {\bibfield  {journal} {\bibinfo  {journal} {Nano Letters}\ }\textbf {\bibinfo
  {volume} {17}},\ \bibinfo {pages} {1637} (\bibinfo {year}
  {2017})}\BibitemShut {NoStop}%
\bibitem [{\citenamefont {Oike}\ \emph {et~al.}(2016)\citenamefont {Oike},
  \citenamefont {Kikkawa}, \citenamefont {Kanazawa}, \citenamefont {Taguchi},
  \citenamefont {Kawasaki}, \citenamefont {Tokura},\ and\ \citenamefont
  {Kagawa}}]{oike_interplay_2016}%
  \BibitemOpen
  \bibfield  {author} {\bibinfo {author} {\bibfnamefont {H.}~\bibnamefont
  {Oike}}, \bibinfo {author} {\bibfnamefont {A.}~\bibnamefont {Kikkawa}},
  \bibinfo {author} {\bibfnamefont {N.}~\bibnamefont {Kanazawa}}, \bibinfo
  {author} {\bibfnamefont {Y.}~\bibnamefont {Taguchi}}, \bibinfo {author}
  {\bibfnamefont {M.}~\bibnamefont {Kawasaki}}, \bibinfo {author}
  {\bibfnamefont {Y.}~\bibnamefont {Tokura}}, \ and\ \bibinfo {author}
  {\bibfnamefont {F.}~\bibnamefont {Kagawa}},\ }\href {\doibase
  10.1038/nphys3506} {\bibfield  {journal} {\bibinfo  {journal} {Nature
  Physics}\ }\textbf {\bibinfo {volume} {12}},\ \bibinfo {pages} {62} (\bibinfo
  {year} {2016})}\BibitemShut {NoStop}%
\bibitem [{\citenamefont {Nakajima}\ \emph {et~al.}(2017)\citenamefont
  {Nakajima}, \citenamefont {Oike}, \citenamefont {Kikkawa}, \citenamefont
  {Gilbert}, \citenamefont {Booth}, \citenamefont {Kakurai}, \citenamefont
  {Taguchi}, \citenamefont {Tokura}, \citenamefont {Kagawa},\ and\
  \citenamefont {Arima}}]{nakajima_skyrmion_2017}%
  \BibitemOpen
  \bibfield  {author} {\bibinfo {author} {\bibfnamefont {T.}~\bibnamefont
  {Nakajima}}, \bibinfo {author} {\bibfnamefont {H.}~\bibnamefont {Oike}},
  \bibinfo {author} {\bibfnamefont {A.}~\bibnamefont {Kikkawa}}, \bibinfo
  {author} {\bibfnamefont {E.~P.}\ \bibnamefont {Gilbert}}, \bibinfo {author}
  {\bibfnamefont {N.}~\bibnamefont {Booth}}, \bibinfo {author} {\bibfnamefont
  {K.}~\bibnamefont {Kakurai}}, \bibinfo {author} {\bibfnamefont
  {Y.}~\bibnamefont {Taguchi}}, \bibinfo {author} {\bibfnamefont
  {Y.}~\bibnamefont {Tokura}}, \bibinfo {author} {\bibfnamefont
  {F.}~\bibnamefont {Kagawa}}, \ and\ \bibinfo {author} {\bibfnamefont {T.-h.}\
  \bibnamefont {Arima}},\ }\href {\doibase 10.1126/sciadv.1602562} {\bibfield
  {journal} {\bibinfo  {journal} {Science Advances}\ }\textbf {\bibinfo
  {volume} {3}},\ \bibinfo {pages} {e1602562} (\bibinfo {year}
  {2017})}\BibitemShut {NoStop}%
\bibitem [{\citenamefont {Kagawa}\ \emph {et~al.}(2017)\citenamefont {Kagawa},
  \citenamefont {Oike}, \citenamefont {Koshibae}, \citenamefont {Kikkawa},
  \citenamefont {Okamura}, \citenamefont {Taguchi}, \citenamefont {Nagaosa},\
  and\ \citenamefont {Tokura}}]{kagawa_current-induced_2017}%
  \BibitemOpen
  \bibfield  {author} {\bibinfo {author} {\bibfnamefont {F.}~\bibnamefont
  {Kagawa}}, \bibinfo {author} {\bibfnamefont {H.}~\bibnamefont {Oike}},
  \bibinfo {author} {\bibfnamefont {W.}~\bibnamefont {Koshibae}}, \bibinfo
  {author} {\bibfnamefont {A.}~\bibnamefont {Kikkawa}}, \bibinfo {author}
  {\bibfnamefont {Y.}~\bibnamefont {Okamura}}, \bibinfo {author} {\bibfnamefont
  {Y.}~\bibnamefont {Taguchi}}, \bibinfo {author} {\bibfnamefont
  {N.}~\bibnamefont {Nagaosa}}, \ and\ \bibinfo {author} {\bibfnamefont
  {Y.}~\bibnamefont {Tokura}},\ }\href {\doibase 10.1038/s41467-017-01353-2}
  {\bibfield  {journal} {\bibinfo  {journal} {Nature Communications}\ }\textbf
  {\bibinfo {volume} {8}},\ \bibinfo {pages} {1332} (\bibinfo {year}
  {2017})}\BibitemShut {NoStop}%
\bibitem [{\citenamefont {Milde}\ \emph {et~al.}(2013)\citenamefont {Milde},
  \citenamefont {Köhler}, \citenamefont {Seidel}, \citenamefont {Eng},
  \citenamefont {Bauer}, \citenamefont {Chacon}, \citenamefont {Kindervater},
  \citenamefont {Mühlbauer}, \citenamefont {Pfleiderer}, \citenamefont
  {Buhrandt}, \citenamefont {Schütte},\ and\ \citenamefont
  {Rosch}}]{milde_unwinding_2013}%
  \BibitemOpen
  \bibfield  {author} {\bibinfo {author} {\bibfnamefont {P.}~\bibnamefont
  {Milde}}, \bibinfo {author} {\bibfnamefont {D.}~\bibnamefont {Köhler}},
  \bibinfo {author} {\bibfnamefont {J.}~\bibnamefont {Seidel}}, \bibinfo
  {author} {\bibfnamefont {L.~M.}\ \bibnamefont {Eng}}, \bibinfo {author}
  {\bibfnamefont {A.}~\bibnamefont {Bauer}}, \bibinfo {author} {\bibfnamefont
  {A.}~\bibnamefont {Chacon}}, \bibinfo {author} {\bibfnamefont
  {J.}~\bibnamefont {Kindervater}}, \bibinfo {author} {\bibfnamefont
  {S.}~\bibnamefont {Mühlbauer}}, \bibinfo {author} {\bibfnamefont
  {C.}~\bibnamefont {Pfleiderer}}, \bibinfo {author} {\bibfnamefont
  {S.}~\bibnamefont {Buhrandt}}, \bibinfo {author} {\bibfnamefont
  {C.}~\bibnamefont {Schütte}}, \ and\ \bibinfo {author} {\bibfnamefont
  {A.}~\bibnamefont {Rosch}},\ }\href {\doibase 10.1126/science.1234657}
  {\bibfield  {journal} {\bibinfo  {journal} {Science}\ }\textbf {\bibinfo
  {volume} {340}},\ \bibinfo {pages} {1076} (\bibinfo {year}
  {2013})}\BibitemShut {NoStop}%
\bibitem [{\citenamefont {M{\"u}nzer}\ \emph {et~al.}(2010)\citenamefont
  {M{\"u}nzer}, \citenamefont {Neubauer}, \citenamefont {Adams}, \citenamefont
  {M{\"u}hlbauer}, \citenamefont {Franz}, \citenamefont {Jonietz},
  \citenamefont {Georgii}, \citenamefont {Böni}, \citenamefont {Pedersen},
  \citenamefont {Schmidt}, \citenamefont {Rosch},\ and\ \citenamefont
  {Pfleiderer}}]{munzer_skyrmion_2010}%
  \BibitemOpen
  \bibfield  {author} {\bibinfo {author} {\bibfnamefont {W.}~\bibnamefont
  {M{\"u}nzer}}, \bibinfo {author} {\bibfnamefont {A.}~\bibnamefont
  {Neubauer}}, \bibinfo {author} {\bibfnamefont {T.}~\bibnamefont {Adams}},
  \bibinfo {author} {\bibfnamefont {S.}~\bibnamefont {M{\"u}hlbauer}}, \bibinfo
  {author} {\bibfnamefont {C.}~\bibnamefont {Franz}}, \bibinfo {author}
  {\bibfnamefont {F.}~\bibnamefont {Jonietz}}, \bibinfo {author} {\bibfnamefont
  {R.}~\bibnamefont {Georgii}}, \bibinfo {author} {\bibfnamefont
  {P.}~\bibnamefont {Böni}}, \bibinfo {author} {\bibfnamefont
  {B.}~\bibnamefont {Pedersen}}, \bibinfo {author} {\bibfnamefont
  {M.}~\bibnamefont {Schmidt}}, \bibinfo {author} {\bibfnamefont
  {A.}~\bibnamefont {Rosch}}, \ and\ \bibinfo {author} {\bibfnamefont
  {C.}~\bibnamefont {Pfleiderer}},\ }\href {\doibase
  10.1103/PhysRevB.81.041203} {\bibfield  {journal} {\bibinfo  {journal}
  {Physical Review B}\ }\textbf {\bibinfo {volume} {81}},\ \bibinfo {pages}
  {041203} (\bibinfo {year} {2010})}\BibitemShut {NoStop}%
\bibitem [{\citenamefont {Okamura}\ \emph {et~al.}(2016)\citenamefont
  {Okamura}, \citenamefont {Kagawa}, \citenamefont {Seki},\ and\ \citenamefont
  {Tokura}}]{okamura_transition_2016}%
  \BibitemOpen
  \bibfield  {author} {\bibinfo {author} {\bibfnamefont {Y.}~\bibnamefont
  {Okamura}}, \bibinfo {author} {\bibfnamefont {F.}~\bibnamefont {Kagawa}},
  \bibinfo {author} {\bibfnamefont {S.}~\bibnamefont {Seki}}, \ and\ \bibinfo
  {author} {\bibfnamefont {Y.}~\bibnamefont {Tokura}},\ }\href {\doibase
  10.1038/ncomms12669} {\bibfield  {journal} {\bibinfo  {journal} {Nature
  Communications}\ }\textbf {\bibinfo {volume} {7}},\ \bibinfo {pages} {12669}
  (\bibinfo {year} {2016})}\BibitemShut {NoStop}%
\bibitem [{\citenamefont {Wild}\ \emph {et~al.}(2017)\citenamefont {Wild},
  \citenamefont {Meier}, \citenamefont {Pöllath}, \citenamefont {Kronseder},
  \citenamefont {Bauer}, \citenamefont {Chacon}, \citenamefont {Halder},
  \citenamefont {Schowalter}, \citenamefont {Rosenauer}, \citenamefont {Zweck},
  \citenamefont {Müller}, \citenamefont {Rosch}, \citenamefont {Pfleiderer},\
  and\ \citenamefont {Back}}]{wild_entropy-limited_2017}%
  \BibitemOpen
  \bibfield  {author} {\bibinfo {author} {\bibfnamefont {J.}~\bibnamefont
  {Wild}}, \bibinfo {author} {\bibfnamefont {T.~N.~G.}\ \bibnamefont {Meier}},
  \bibinfo {author} {\bibfnamefont {S.}~\bibnamefont {Pöllath}}, \bibinfo
  {author} {\bibfnamefont {M.}~\bibnamefont {Kronseder}}, \bibinfo {author}
  {\bibfnamefont {A.}~\bibnamefont {Bauer}}, \bibinfo {author} {\bibfnamefont
  {A.}~\bibnamefont {Chacon}}, \bibinfo {author} {\bibfnamefont
  {M.}~\bibnamefont {Halder}}, \bibinfo {author} {\bibfnamefont
  {M.}~\bibnamefont {Schowalter}}, \bibinfo {author} {\bibfnamefont
  {A.}~\bibnamefont {Rosenauer}}, \bibinfo {author} {\bibfnamefont
  {J.}~\bibnamefont {Zweck}}, \bibinfo {author} {\bibfnamefont
  {J.}~\bibnamefont {Müller}}, \bibinfo {author} {\bibfnamefont
  {A.}~\bibnamefont {Rosch}}, \bibinfo {author} {\bibfnamefont
  {C.}~\bibnamefont {Pfleiderer}}, \ and\ \bibinfo {author} {\bibfnamefont
  {C.~H.}\ \bibnamefont {Back}},\ }\href {\doibase 10.1126/sciadv.1701704}
  {\bibfield  {journal} {\bibinfo  {journal} {Science Advances}\ }\textbf
  {\bibinfo {volume} {3}},\ \bibinfo {pages} {e1701704} (\bibinfo {year}
  {2017})}\BibitemShut {NoStop}%
\bibitem [{\citenamefont {Parkin}\ \emph {et~al.}(2008)\citenamefont {Parkin},
  \citenamefont {Hayashi},\ and\ \citenamefont
  {Thomas}}]{parkin_magnetic_2008}%
  \BibitemOpen
  \bibfield  {author} {\bibinfo {author} {\bibfnamefont {S.~S.~P.}\
  \bibnamefont {Parkin}}, \bibinfo {author} {\bibfnamefont {M.}~\bibnamefont
  {Hayashi}}, \ and\ \bibinfo {author} {\bibfnamefont {L.}~\bibnamefont
  {Thomas}},\ }\href {\doibase 10.1126/science.1145799} {\bibfield  {journal}
  {\bibinfo  {journal} {Science}\ }\textbf {\bibinfo {volume} {320}},\ \bibinfo
  {pages} {190} (\bibinfo {year} {2008})}\BibitemShut {NoStop}%
\bibitem [{\citenamefont {Kolton}\ \emph {et~al.}(2009)\citenamefont {Kolton},
  \citenamefont {Rosso}, \citenamefont {Giamarchi},\ and\ \citenamefont
  {Krauth}}]{kolton_creep_2009}%
  \BibitemOpen
  \bibfield  {author} {\bibinfo {author} {\bibfnamefont {A.~B.}\ \bibnamefont
  {Kolton}}, \bibinfo {author} {\bibfnamefont {A.}~\bibnamefont {Rosso}},
  \bibinfo {author} {\bibfnamefont {T.}~\bibnamefont {Giamarchi}}, \ and\
  \bibinfo {author} {\bibfnamefont {W.}~\bibnamefont {Krauth}},\ }\href
  {\doibase 10.1103/PhysRevB.79.184207} {\bibfield  {journal} {\bibinfo
  {journal} {Physical Review B}\ }\textbf {\bibinfo {volume} {79}},\ \bibinfo
  {pages} {184207} (\bibinfo {year} {2009})}\BibitemShut {NoStop}%
\bibitem [{\citenamefont {Chauve}\ \emph {et~al.}(2000)\citenamefont {Chauve},
  \citenamefont {Giamarchi},\ and\ \citenamefont
  {Le~Doussal}}]{chauve_creep_2000}%
  \BibitemOpen
  \bibfield  {author} {\bibinfo {author} {\bibfnamefont {P.}~\bibnamefont
  {Chauve}}, \bibinfo {author} {\bibfnamefont {T.}~\bibnamefont {Giamarchi}}, \
  and\ \bibinfo {author} {\bibfnamefont {P.}~\bibnamefont {Le~Doussal}},\
  }\href {\doibase 10.1103/PhysRevB.62.6241} {\bibfield  {journal} {\bibinfo
  {journal} {Physical Review B}\ }\textbf {\bibinfo {volume} {62}},\ \bibinfo
  {pages} {6241} (\bibinfo {year} {2000})}\BibitemShut {NoStop}%
\bibitem [{\citenamefont {Kleemann}\ \emph {et~al.}(2007)\citenamefont
  {Kleemann}, \citenamefont {Rhensius}, \citenamefont {Petracic}, \citenamefont
  {Ferré}, \citenamefont {Jamet},\ and\ \citenamefont
  {Bernas}}]{kleemann_modes_2007}%
  \BibitemOpen
  \bibfield  {author} {\bibinfo {author} {\bibfnamefont {W.}~\bibnamefont
  {Kleemann}}, \bibinfo {author} {\bibfnamefont {J.}~\bibnamefont {Rhensius}},
  \bibinfo {author} {\bibfnamefont {O.}~\bibnamefont {Petracic}}, \bibinfo
  {author} {\bibfnamefont {J.}~\bibnamefont {Ferré}}, \bibinfo {author}
  {\bibfnamefont {J.~P.}\ \bibnamefont {Jamet}}, \ and\ \bibinfo {author}
  {\bibfnamefont {H.}~\bibnamefont {Bernas}},\ }\href {\doibase
  10.1103/PhysRevLett.99.097203} {\bibfield  {journal} {\bibinfo  {journal}
  {Physical Review Letters}\ }\textbf {\bibinfo {volume} {99}},\ \bibinfo
  {pages} {097203} (\bibinfo {year} {2007})}\BibitemShut {NoStop}%
\bibitem [{\citenamefont {\v{S}tefan\v{c}i\v{c}}\ \emph
  {et~al.}(2018)\citenamefont {\v{S}tefan\v{c}i\v{c}}, \citenamefont {Moody},
  \citenamefont {Hicken}, \citenamefont {Birch}, \citenamefont {Balakrishnan},
  \citenamefont {Barnett}, \citenamefont {Crisanti}, \citenamefont {Evans},
  \citenamefont {Holt}, \citenamefont {Franke}, \citenamefont {Hatton},
  \citenamefont {Huddart}, \citenamefont {Lees}, \citenamefont {Pratt},
  \citenamefont {Tang}, \citenamefont {Wilson}, \citenamefont {Xiao},\ and\
  \citenamefont {Lancaster}}]{stefancic_origin_2018}%
  \BibitemOpen
  \bibfield  {author} {\bibinfo {author} {\bibfnamefont {A.}~\bibnamefont
  {\v{S}tefan\v{c}i\v{c}}}, \bibinfo {author} {\bibfnamefont {S.~H.}\
  \bibnamefont {Moody}}, \bibinfo {author} {\bibfnamefont {T.~J.}\ \bibnamefont
  {Hicken}}, \bibinfo {author} {\bibfnamefont {M.~T.}\ \bibnamefont {Birch}},
  \bibinfo {author} {\bibfnamefont {G.}~\bibnamefont {Balakrishnan}}, \bibinfo
  {author} {\bibfnamefont {S.~A.}\ \bibnamefont {Barnett}}, \bibinfo {author}
  {\bibfnamefont {M.}~\bibnamefont {Crisanti}}, \bibinfo {author}
  {\bibfnamefont {J.~S.~O.}\ \bibnamefont {Evans}}, \bibinfo {author}
  {\bibfnamefont {S.~J.~R.}\ \bibnamefont {Holt}}, \bibinfo {author}
  {\bibfnamefont {K.~J.~A.}\ \bibnamefont {Franke}}, \bibinfo {author}
  {\bibfnamefont {P.~D.}\ \bibnamefont {Hatton}}, \bibinfo {author}
  {\bibfnamefont {B.~M.}\ \bibnamefont {Huddart}}, \bibinfo {author}
  {\bibfnamefont {M.~R.}\ \bibnamefont {Lees}}, \bibinfo {author}
  {\bibfnamefont {F.~L.}\ \bibnamefont {Pratt}}, \bibinfo {author}
  {\bibfnamefont {C.~C.}\ \bibnamefont {Tang}}, \bibinfo {author}
  {\bibfnamefont {M.~N.}\ \bibnamefont {Wilson}}, \bibinfo {author}
  {\bibfnamefont {F.}~\bibnamefont {Xiao}}, \ and\ \bibinfo {author}
  {\bibfnamefont {T.}~\bibnamefont {Lancaster}},\ }\href {\doibase
  10.1103/PhysRevMaterials.2.111402} {\bibfield  {journal} {\bibinfo  {journal}
  {Physical Review Materials}\ }\textbf {\bibinfo {volume} {2}},\ \bibinfo
  {pages} {111402} (\bibinfo {year} {2018})}\BibitemShut {NoStop}%
\bibitem [{\citenamefont {Okamura}\ \emph {et~al.}(2013)\citenamefont
  {Okamura}, \citenamefont {Kagawa}, \citenamefont {Mochizuki}, \citenamefont
  {Kubota}, \citenamefont {Seki}, \citenamefont {Ishiwata}, \citenamefont
  {Kawasaki}, \citenamefont {Onose},\ and\ \citenamefont
  {Tokura}}]{okamura_microwave_2013}%
  \BibitemOpen
  \bibfield  {author} {\bibinfo {author} {\bibfnamefont {Y.}~\bibnamefont
  {Okamura}}, \bibinfo {author} {\bibfnamefont {F.}~\bibnamefont {Kagawa}},
  \bibinfo {author} {\bibfnamefont {M.}~\bibnamefont {Mochizuki}}, \bibinfo
  {author} {\bibfnamefont {M.}~\bibnamefont {Kubota}}, \bibinfo {author}
  {\bibfnamefont {S.}~\bibnamefont {Seki}}, \bibinfo {author} {\bibfnamefont
  {S.}~\bibnamefont {Ishiwata}}, \bibinfo {author} {\bibfnamefont
  {M.}~\bibnamefont {Kawasaki}}, \bibinfo {author} {\bibfnamefont
  {Y.}~\bibnamefont {Onose}}, \ and\ \bibinfo {author} {\bibfnamefont
  {Y.}~\bibnamefont {Tokura}},\ }\href {\doibase 10.1038/ncomms3391} {\bibfield
   {journal} {\bibinfo  {journal} {Nature Communications}\ }\textbf {\bibinfo
  {volume} {4}},\ \bibinfo {pages} {2391} (\bibinfo {year} {2013})}\BibitemShut
  {NoStop}%
\bibitem [{\citenamefont {Garst}\ \emph {et~al.}(2017)\citenamefont {Garst},
  \citenamefont {Waizner},\ and\ \citenamefont
  {Grundler}}]{garst_collective_2017}%
  \BibitemOpen
  \bibfield  {author} {\bibinfo {author} {\bibfnamefont {M.}~\bibnamefont
  {Garst}}, \bibinfo {author} {\bibfnamefont {J.}~\bibnamefont {Waizner}}, \
  and\ \bibinfo {author} {\bibfnamefont {D.}~\bibnamefont {Grundler}},\ }\href
  {\doibase 10.1088/1361-6463/aa7573} {\bibfield  {journal} {\bibinfo
  {journal} {Journal of Physics D: Applied Physics}\ }\textbf {\bibinfo
  {volume} {50}},\ \bibinfo {pages} {293002} (\bibinfo {year}
  {2017})}\BibitemShut {NoStop}%
\bibitem [{\citenamefont {Mochizuki}\ and\ \citenamefont
  {Seki}(2015)}]{mochizuki_dynamical_2015}%
  \BibitemOpen
  \bibfield  {author} {\bibinfo {author} {\bibfnamefont {M.}~\bibnamefont
  {Mochizuki}}\ and\ \bibinfo {author} {\bibfnamefont {S.}~\bibnamefont
  {Seki}},\ }\href {\doibase 10.1088/0953-8984/27/50/503001} {\bibfield
  {journal} {\bibinfo  {journal} {Journal of Physics: Condensed Matter}\
  }\textbf {\bibinfo {volume} {27}},\ \bibinfo {pages} {503001} (\bibinfo
  {year} {2015})}\BibitemShut {NoStop}%
\bibitem [{\citenamefont {Onose}\ \emph {et~al.}(2012)\citenamefont {Onose},
  \citenamefont {Okamura}, \citenamefont {Seki}, \citenamefont {Ishiwata},\
  and\ \citenamefont {Tokura}}]{onose_observation_2012}%
  \BibitemOpen
  \bibfield  {author} {\bibinfo {author} {\bibfnamefont {Y.}~\bibnamefont
  {Onose}}, \bibinfo {author} {\bibfnamefont {Y.}~\bibnamefont {Okamura}},
  \bibinfo {author} {\bibfnamefont {S.}~\bibnamefont {Seki}}, \bibinfo {author}
  {\bibfnamefont {S.}~\bibnamefont {Ishiwata}}, \ and\ \bibinfo {author}
  {\bibfnamefont {Y.}~\bibnamefont {Tokura}},\ }\href {\doibase
  10.1103/PhysRevLett.109.037603} {\bibfield  {journal} {\bibinfo  {journal}
  {Physical Review Letters}\ }\textbf {\bibinfo {volume} {109}},\ \bibinfo
  {pages} {037603} (\bibinfo {year} {2012})}\BibitemShut {NoStop}%
\bibitem [{\citenamefont {Bannenberg}\ \emph {et~al.}(2016)\citenamefont
  {Bannenberg}, \citenamefont {Lefering}, \citenamefont {Kakurai},
  \citenamefont {Onose}, \citenamefont {Endoh}, \citenamefont {Tokura},\ and\
  \citenamefont {Pappas}}]{bannenberg_magnetic_2016}%
  \BibitemOpen
  \bibfield  {author} {\bibinfo {author} {\bibfnamefont {L.~J.}\ \bibnamefont
  {Bannenberg}}, \bibinfo {author} {\bibfnamefont {A.~J.~E.}\ \bibnamefont
  {Lefering}}, \bibinfo {author} {\bibfnamefont {K.}~\bibnamefont {Kakurai}},
  \bibinfo {author} {\bibfnamefont {Y.}~\bibnamefont {Onose}}, \bibinfo
  {author} {\bibfnamefont {Y.}~\bibnamefont {Endoh}}, \bibinfo {author}
  {\bibfnamefont {Y.}~\bibnamefont {Tokura}}, \ and\ \bibinfo {author}
  {\bibfnamefont {C.}~\bibnamefont {Pappas}},\ }\href {\doibase
  10.1103/PhysRevB.94.134433} {\bibfield  {journal} {\bibinfo  {journal}
  {Physical Review B}\ }\textbf {\bibinfo {volume} {94}},\ \bibinfo {pages}
  {134433} (\bibinfo {year} {2016})}\BibitemShut {NoStop}%
\end{thebibliography}%

\end{document}